\documentclass[a4paper,12pt]{article}

\usepackage[latin1]{inputenc}
\usepackage[american]{babel}

\usepackage{euscript}
\usepackage{amssymb}
\usepackage{amsfonts}
\usepackage{amsbsy}
\usepackage{amsmath}
\usepackage{epsfig}
\usepackage{hyperref}

 \usepackage{a4wide}
 \usepackage{graphics}
 \usepackage{subfigure}
 \usepackage{fancyhdr}

\newcommand{\bee}{\begin{equation}}
\newcommand{\ee}{\end{equation}}
\newcommand{\bea}{\begin{eqnarray}}
\newcommand{\eea}{\end{eqnarray}}

\newcommand{\NN}{\mathcal{N}}
\newcommand{\FF}{\mathcal{F}}

\newcommand{\Ss}{\mathcal{S}}

\newcommand{\OO}{\mathcal{O}}
\newcommand{\Rbb}{\mathbb{R}}
\newcommand{\Zbb}{\mathbb{Z}}
\newcommand{\Cbb}{\mathbb{C}}

\newcommand{\nn}{\nonumber}

\newcommand{\re}{\mbox{Re}}
\newcommand{\im}{\mbox{Im}}

\newcommand{\Vol}{\mbox{Vol}}
\newcommand{\vol}{\hat{\mathcal{V}}}

\def\bF{\bar{F}}

\newcommand{\der}{\partial}
\newcommand{\e}{\mathrm{e}}

\newcommand{\dd}{\mathrm{d}}

\begin{document}

\date{}
\rightline{HD-THEP-07-28}
\rightline{SISSA 74/2007/EP}
\pagestyle{empty}
\begin{center}
{\Large\bf Type IIA/M-theory Moduli fixing\\ in a Class of Orientifold Models}
\\[2.1em]

\bigskip

{\large Giuseppe Milanesi $^{a,c,d,\dagger}$, Roberto Valandro$^{b,c,d,*}$}\\

\null

\noindent 
{\it $^a$  Institut f\"ur Theoretische Physik,\\
ETH Z\"urich,
CH-8093 Z\"urich, Switzerland  }
\\[2.1ex]
{\it $^b$ Institut f\"ur Theoretische Physik, Universit\"at Heidelberg,\\
 Philosophenweg 16 und 19, D-69120 Heidelberg, Germany}
\\[2.1ex]
  {\it $^c$ Scuola Internazionale Superiore di Studi
Avanzati, \\
Via Beirut 2-4, 34014 Trieste, Italy }\\[2.1ex]
{\it $^d$ Istituto Nazionale di Fisica Nucleare, Sez. di Trieste}\\[2.1ex]
 \vfill

\end{center}

\begin{abstract}
We present the study of type II A flux vacua and their M-theory duals for compactification on a class of Calabi-Yau orientifolds. The K\"ahler potential is derived from toroidal compactifications and the superpotential contains a contribution from non-Abelian gauge degrees of freedoms. We obtain complete stabilisation of the moduli. We found one supersymmetric minimum and several non supersymmetric ones. Consistency of the analysis constrains the parameters of the models in a finite region containing a finite, although very large, number of flux vacua. From the M-theory side, we found some differences in the distributions of the physical quantities with respect to the M-theory ensemble studied by Acharya {\it et al} \cite{Acharya:2005ez}. In particular, it is easier to find small supersymmetry breaking scale.
\end{abstract}

\vfill 
\leftline{$^\dagger$ milanesi@itp.phys.ethz.ch}
\leftline{ $^*$ r.valandro@thphys.uni-heidelberg.de}
\newpage

\pagestyle{plain}

%

\section{Introduction}

String/M-theory theory has long held the promise to provide us with a complete and final description of the laws of physics in our universe. In its original formulation it is a ten(eleven) dimensional theory, while the low energy physics is described by a four dimensional theory.
One approach to reduce String/M-theory from ten(eleven) to four spacetime dimensions is the so called {\it process of compactification}. It consists in studying the theory on a geometric background of the form $M^{3,1}\times X$. $M^{3,1}$ is identified with our spacetime, while the manifold $X$ is chosen to be {\it small} and {\it compact}, such that the six(seven) additional dimensions are not detectable in experiments. 

The process of compactification introduces a high amount of ambiguity, as String/M-theory allows many different choices of $X$.
To get the effective four dimensional theory, one should integrate out the massive string states\cite{Polchinski:1998rq}, together with the massive Kaluza-Klein (KK)\cite{Kaluza:1921tu,Klein:1926tv} modes appearing in the process of compactification.
The structure of the obtained four dimensional theory strongly depends on the chosen internal manifold $X$. The properties of $X$ determine the amount of preserved supersymmetry and the surviving gauge group of the lower dimensional effective theory. Usually one requires $X$ to preserve some supercharges, both for phenomenological reasons and because String/M-theory on supersymmetric background is under much better control than on non-supersymmetric ones. This requirement is actually translated into a geometric condition on the compact manifold: it must have reduced holonomy. In particular in many cases this implies the internal manifold to be a Calabi-Yau (CY), {\it i.e} a six dimensional compact manifold with $SU(3)$ holonomy. After compactification and reduction to the four dimensional theory, one would like at least to obtain a realistic spectrum. But here one encounters
one of the main problems in compactification: the presence of {\it moduli}. These are parameters that label continuous degeneracies of the metric of a consistent choice of a compact manifold $X$ corresponding to fluctuations of its size and shape. They can generically take arbitrary values. In four dimensions, they 
appear as massless neutral scalar fields. These scalars are not present in our world and one should find a mechanism to generate a potential for them, in such a way that they acquire a mass and are not dynamical in the low energy action. 
Moreover, the low energy masses and coupling constants are functions of the moduli.

In order to introduce a potential that stabilises the moduli, one should add some new ingredients to the compactification. One of them, largely studied in the recent years, is the introduction of non-zero fluxes threading nontrivial cycles of the compact manifold. Each of the limits of M-theory mentioned above has certain p-form gauge fields, which are sourced by elementary branes. Background values for their field strength can actually stabilise the moduli. This is because, their contribution to the total energy will depend on the moduli controlling the size of the cycles that the fluxes are threading. If the generated potential is sufficiently general, minimising it will stabilise the moduli to fixed values. Some beautiful recent reviews on flux compactifications are \cite{Grana:2005jc,Douglas:2006es,Blumenhagen:2006ci}.

The fluxes are subject to a Dirac-like quantisation condition. Hence they take discrete values, that add to the other discrete parameters parametrising the compactification data, such as for instance the brane charges. The four dimensional effective moduli potential depends on these discrete data. Varying them we get an ensemble of effective four dimensional potentials. Minimising the potential for different values of the parameters we obtain a set of vacua. Putting all together, one gets an huge number of lower dimensional string groundstates (vacua). The set of all these four dimensional constructions is called ``the Landscape''.

\

In this paper we will consider a particular subset of the Landscape: Type IIA compactification with fluxes. In \cite{DeWolfe:2005uu} it was found that turning on all possible RR and NSNS fluxes, fixes all the CY geometric moduli at the classical level. One could try to go to the M-theory dual of this compactification, but some fluxes transform to geometry and the compactification manifold is no more a special holonomy manifold. If one starts from M-theory compactification on a $G_2$ manifold with fluxes turned on and dualises to type IIA, one gets that the only non vanishing fluxes are the 3-form and 4-form: neither in the M-theory nor in the Type IIA  setting this is enough to fix all the geometric moduli \cite{Beasley:2002db}. In \cite{Acharya:2002kv} it was found a mechanism to fix all the geometric moduli in M-theory compactification with fluxes by turning on non-Abelian degrees of freedom coming from the gauge theory living on ADE singularities of the compact manifold. These give a constant contribution to the superpotential that stabilises all the moduli. 

%
We consider the same contribution to the superpotential in Type IIA compactifications on CY. In this case, the degrees of freedom responsible for the constant term in the superpotential come from non-Abelian background gauge fields living on D6-branes. We will consider models with $H_3$ and $F_4$ turned on and $m$, $F_2$, $F_6$ vanishing: this assumption simplifies the scalar potential with respect to the generic case with all the fluxes turned on. 
The potential is proven to be formally equal to the M-theory dual, once identified the moduli and the fluxes in the two theories. So the mechanism studied in \cite{Acharya:2002kv} works for this class of Type IIA vacua, and we are assured that all the moduli are fixed.

Given an explicit formula for the K\"ahler potential and the superpotential, one can compute the scalar potential. We will consider a class of models for which the K\"ahler potential is suggested by studying toroidal orbifold compactifications. We minimise the corresponding potential and find all the extremal points. In particular we find a non-supersymmetric $AdS$ absolute minimum that could become $dS$ after uplifting. The results are valid both for Type IIA flux vacua and for the M-theory duals. We will study the consistency of the solutions with the approximations done and compare the results with the M-theory flux vacua of \cite{Acharya:2005ez}, that were found using a different K\"ahler potential.

\

The paper is organised as follows. In the next Section we describe M-theory and type IIA vacua, focusing on the potential generated by fluxes and how the two sets of vacua are related.
In Section 3 we discuss Type IIA flux compactification on orbifolds and the K\"ahler potential for the untwisted moduli. Using them, we will get an explicit form of the scalar potential. In Section 4  we study in detail the potential for a particular set of orbifold models. We obtain all the minima of the potential and discuss the physical properties of the two most interesting vacua. We also briefly consider the same potential in the dual M-theory picture. In Section 5 we present a note on M-theory vacua without flux and final in Section 6 we give our conclusions. In the Appendix detailed calculation referring to Section 4 are reported.

\section{M-theory and Type IIA Flux Vacua}

\subsection{M-theory Flux Vacua}

In this section we will review the M-theory compactification with fluxes.

M-theory is locally supersymmetric and is described at low energy by the eleven dimensional supergravity. Its action is given by:
\bee
 S = \frac{1}{2\kappa_{11}^2}\left[\int d^{11}x  \sqrt{-g}R -\int \left(\frac{1}{2} G\wedge \ast G - \frac{1}{6} C\wedge G\wedge G\right)\right]
\ee
The bosonic fields are the eleven dimensional metric and a 3-form $C$, whose field strength is $G=dC$.

To obtain a four dimensional theory, we have to compactify on a seven dimensional manifold $X$. Requiring $\NN=1$ supersymmetry in four dimension poses constraints on the holonomy group of $X$, {\it i.e.} it should be $G_2$. A central point concerning such $G_2$ compactification is that, if $X$ is smooth, the four dimensional physics contains at most Abelian gauge group and no light charged fermions.  These features arise in the effective theory if we compactify on a singular $G_2$ holonomy manifold \cite{Acharya:2000gb,Acharya:2001gy}. In particular if $X$ admits a three dimensional locus $Q$ of ADE singularities, the low energy theory contains a SYM theory on $M^{3,1}\times Q$ \cite{Acharya:2000gb}.

The $G_2$ holonomy allows for exactly one covariantly constant spinor which can be used to define a real, harmonic and covariantly constant 3-form $\Phi$. The moduli space of the metric has dimension $b^3(X)=\dim H^3(X,\Rbb)$ and can be parametrised by expanding $\Phi$ into the basis of harmonic 3-forms $\phi_i$:
\bee
 \Phi = s^i(x) \phi_i
\ee
The scalars $s^i$ are combined to the ones coming from the KK expansion of the 3-form potential $C$:
\bee
 C = t^i(x) \phi_i
\ee 
The complex scalars $z^i=t^i+i s^i$ form the bosonic components of $b^3(X)$ chiral multiplet.

The four dimensional theory is an $\NN=1$ supergravity. Its action is completely determined in terms of the K\"ahler potential $K$, the superpotential $W$ and the gauge kinetic functions $f$.

The K\"ahler potential is given by:
\bea\label{MthKpot}
 K = -3 \ln \left(V_X \right) &\mbox{ with }& V_X = \frac{1}{7}\int_X \Phi\wedge \ast \Phi
\eea
$V_X$ is the compactification volume in Planck units, as $V_X = \Vol X \, m_P^{7}$, where $m_P$ is the eleven dimensional Planck mass. $V_X$ has to be a homogeneous function of the $s^i$ of degree $7/3$ \cite{Beasley:2002db}.

When a 4-form flux $G$ is turned on, it induces the superpotential \cite{Beasley:2002db}:
\bee\label{MthSuperpot}
 W_0 = \int_X \left(\frac{1}{2}C+i\Phi\right)\wedge G 
\ee
This superpotential does not fix the moduli, as the induced potential is positive definite and runs down to zero at infinite volume. However, \eqref{MthSuperpot} does not include possible contributions coming from the gauge theory living on the ADE-locus $Q$. 
Non-Abelian flux for these degrees of freedom actually gives an additional contribution to the superpotential which stabilise all the moduli \cite{Acharya:2002kv}. It works if the submanifold $Q$ admits a complex, non-real Chern-Simons invariant \cite{Acharya:2002kv}. This is the case if, for example, $Q$ is a hyperbolic manifold. The final superpotential is:
\bee\label{Mthsuppot}
 W = W_0+ (c_1+ic_2) = \sum_{j=1}^{b^3} z_j N_j + c_1 + ic_2
\ee
The expression for $W_0$ is given by expanding $G$ on the basis of harmonic 4-forms that are dual to the $\phi_i$'s on $X$, while
$c=c_1+ic_2$ is the Chern-Simons contribution to the superpotential. In general the constant $c$ is complex. In particular the real part is only well defined modulo 1 in appropriate units and is essentially the more familiar real Chern-Simons invariant. Its imaginary part however can in general take any possibly large real number \cite{Acharya:2002kv}.

\

To write an explicit form for the scalar potential, one needs the exact form of the K\"ahler potential, and in particular of the volume function $V_X$. Unfortunately, unlike the CY case, where the volume function is always a third order homogeneous polynomial in the K\"ahler moduli, no strong constraints on $V_X$ are known for $G_2$ holonomy manifolds. The only condition is that $V_X$ must be a homogeneous function of degree $7/3$ and that the second derivative of $K$ must be positive definite (as it gives the kinetic energies of the moduli). In general it is difficult to find simple candidate volume functions satisfying these constraints. In \cite{Acharya:2005ez} the authors suggested a simple but general formula for $V_X$:
\bea\label{VxMth}
 V_X=\prod_{i=1}^{b^3}(s_i)^{a_i} &{\rm with }& \sum_{i=1}^{b^3}a_i = \frac{7}{3}
\eea
If $a_i>0$ $\forall i$, this gives a positive metric on the moduli space \cite{Acharya:2005ez}.

One is then able to compute the scalar potential, using the standard form:
\bee\label{effpot}
 V = e^K \left(g^{i\bar{j}} D_iW \overline{D_jW} - 3|W|^2\right)
\ee
Here the covariant derivative is given by $D_i W = \partial_i W +\partial_iK\, W$.

This potential was studied in details in \cite{Acharya:2005ez} and all its extremal points were found. The vacua are labelled by $(N_i,\sigma_i)$ ($i=1,...,b^3$), where $N_i$ are the fluxes and $\sigma_i=\pm 1$. Putting all $\sigma_i=+1$ gives a supersymmetric $AdS$ vacuum. The other $2^{b^3}-1$ choices correspond to nonsupersymmetric vacua. Not all of these vacua exist within the supergravity approximation; however an exponentially (in $b^3$) large number survive. In \cite{Acharya:2005ez} it was also shown that all de Sitter vacua are classically unstable, while a large number of non-supersymmetric $AdS$ vacua are metastable.

\subsection{Type IIA Flux Vacua}
To derive a four dimensional description of the Type IIA orientifold vacua with fluxes, one reduces to four dimensions the ten dimensional action of Type IIA massive supergravity \cite{Grimm:2004ua}. In string frame it is given by:
\bea
 S &=& \frac{1}{2\kappa_{10}^2}\int d^{10}x \sqrt{-g} \left(e^{-2\phi}(R+4(\partial_M\phi)^2-\frac{1}{2}|H|^2) - 
		(|\tilde{F}_2|^2 + |\tilde{F}_4|^2+m^2)\right) \nn\\\nn\\
	&& + S_{CS} \:\:+ S_{loc}
\eea
with $2\kappa_{10}^2=(2\pi)^{7}\alpha'^{4}$. 

The fields involved are the metric $g$ , the dilaton $\phi$, the NSNS field strength $H$ (with potential $B$) and the RR field strengths: the zero form $F_0=m$ which is not dynamical, the 2-form $\tilde{F}_2$ and the 4-form $\tilde{F}_4$ (with potentials $C_1$ and  $C_3$).
The physical field strengths are:
\bea
 \tilde{F}_2&=&dC_1+mB\\
 \tilde{F}_4&=&dC_3-\frac{m}{2}B\wedge B - C_1 \wedge H_3
\eea
$S_{CS}$ is the Chern-Simons term and $S_{loc}$ is the contribution of the localised sources included in the compactification.

Here we consider the compactification on a CY 3-fold $Y$, with orientifold O6-planes. In Type IIA the complex structure moduli are promoted to quaternionic multiplets by combining them with the RR axions. The expansion of $\Omega$ and $C_3$ on the basis of harmonic 3-forms is given by:
\bea\label{OmegaC3exp}
 \Omega = Z^{\hat{K}}\alpha_{\hat{K}} - \FF_{\hat{K}} \beta^{\hat{K}}
  && C_3 = \xi^{\hat{K}}\alpha_{\hat{K}} - \tilde{\xi}_{\hat{K}} \beta^{\hat{K}}
\eea
where $\{\alpha_{\hat{K}},\beta^{\hat{K}}\}$ are a basis in $H^3(X)$. The $Z^{\hat{K}}$ ($\hat{K}=0,...,h^{2,1}$) are projective coordinates on the complex structure moduli space (where we can take $z^K=Z^K/Z^0$ with $K\not= 0$ as local coordinates), while $\FF_{\hat{K}}$ are functions of them. From $C_3$, we get $h^{2,1}+1$ complex axions. The axions coming from $\xi^0,\tilde{\xi}_0$ join the axion-dilaton, while the other $h^{2,1}$ quaternionise the $z^K$. The orientifold projection cuts this quaternionic space and the K\"ahler potential is changed sensitively.  

The orientifold projection is given by the operator $\OO=\Omega_p (-1)^{F_L} \sigma$, where $\sigma$ is an antiholomorphic involution of the CY. It acts on the forms $J$ and $\Omega$ as:
\bea
 \sigma^\ast J = - J, && \sigma^\ast\Omega=e^{2i\theta}\bar{\Omega}
\eea
with $\theta$ some arbitrary phase. The orientifold involution splits $H^3=H^3_+ + H^3_-$. Each of these eigenspaces is of real dimension $h^{2,1}+1$. We split the basis for $H^3$ into a set of even forms $\{\alpha_k,\beta^\lambda\}$ and a set of odd forms $\{\alpha_\lambda,\beta^k\}$; here $k=0,...,\tilde{h}$ while $\lambda=\tilde{h}+1,...,h^{2,1}$. Then the orientifold projections requires (taking $\theta=0$):
\bee
 \im Z^k = \re \FF_k = \re Z^\lambda = \im \FF_\lambda =0
\ee
Half of these conditions are constraints on the moduli, while the other half follow automatically for a space admitting the antiholomorphic involution $\sigma$ \cite{Grimm:2004ua}. We see that for each complex $z^K$, only one real component survives the projection. The condition that $C_3$ must be even under $\sigma$ truncates the space of axions in half to $\xi^k,\tilde{\xi}_\lambda$. In addition, the orientifold projects in the dilaton and one of $\xi^0$ and $\tilde{\xi}_0$. So from each hypermultiplet, we get a single chiral multiplet, whose scalar components are the real or imaginary part of the complex structure modulus, and a RR axion.

We can summarise the surviving hypermultiplet moduli in terms of the object
\bee
 \Omega_c = C_3 + 2 i \re (\mathcal{C}\Omega)
\ee
Here, $\mathcal{C}$ is a compensator which incorporates the dilaton dependence via
\bea
 \mathcal{C} = e^{-D+K_C/2}, && e^D = \sqrt{8}e^{\phi+K_k/2}
\eea
where $K_C$ and $K_k$ are the K\"ahler potential for the metric on the CY moduli space \cite{Candelas:1990pi}. One should think of $e^D$ as the four dimensional dilaton.
The surviving moduli are then the expansion of $\Omega_c$ in a basis for $H^3_+$:
\bea\label{Nkdef}
 n^k &=& \frac{1}{2}\int_Y \Omega_c \wedge \beta^k = \frac{1}{2}\xi^k + i\re(\mathcal{C} Z^k)\\
 T_\lambda &=& i \int_Y \Omega_c \wedge \alpha_\lambda = i \tilde{\xi}_\lambda - 2 \re (\mathcal{C} \FF_\lambda)\:.
\eea
The K\"ahler potential which governs the metric on the CY orientifold  complex structure moduli space takes the form:
\bee
 K_Q = - 2 \ln \left(2 \int_Y \re (\mathcal{C}\Omega) \wedge \ast \re(\mathcal{C}\Omega)\right) = -2\ln \left(\im(\mathcal{C} Z^\lambda)\re(\mathcal{C}\FF_\lambda) - \re(\mathcal{C}Z^k)\im(\mathcal{C}\FF_k)\right)
\ee

The K\"ahler potential for the K\"ahler moduli remains of the same form as without orientifold:
\bee\label{KkIIA}
 K_k = - \ln \left(\frac{4}{3} \int_X J\wedge J\wedge J\right)
\ee
The only difference is that only odd fluctuations of $J$ survive. Actually, $J$ is complexified by $B$ in defining:
\bee
 J_c = B + i J
\ee
Since $J_c$ is odd under the orientifold projection, it is expanded on a basis $\omega_a$ of $h^{1,1}_-$ odd harmonic forms:
\bea
 J_c= t^a \omega_a, && t^a= b^a +iv^a
\eea

\

We can now turn on the fluxes which are projected in by the orientifold. It turns out that $H$ and $\tilde{F}_2$ must be odd under the anti-holomorphic involution $\sigma$, while $\tilde{F}_4$ should be even. Se we can expand the fluxes as:
\bea
 H^f=q^\lambda \alpha_\lambda-p_k \beta^k, & F_2^f = - m^a \omega_a, & F_4^f= e_a\tilde{\omega}^a
\eea
where $\tilde{\omega}^a$ are the 4-forms dual of the $\omega_a$. Since the volume form is odd, while the former are even the second are odd. There are in addition two parameters $m$ and $e_0$ parametrising the $F_0$ and $F_6$ fluxes on $Y$.

The background fluxes contribute to the total D6 charge, together with the orientifold O6-plane. Actually the Bianchi identity for $F_2$ is given by
\bee
 d\tilde{F}_2 =  m H -  \mu_6 \delta_3
\ee
where $\delta_3$ is the Poincar\'e dual three-form of the 3-cycle wrapped by the O6-plane. Integrating this equation over any 3-cycle produces a cancellation condition between the combination $mH_3$ of the RR 0-form flux and the NSNS 3-form flux, and the background O6-plane charge. Adding D6-branes also would contribute. This is the analogue of the effective D3 charge of the 3-form fluxes in Type IIB. In Type IIA there are other RR fluxes that are not constrained.

\

The $\NN=1$ potential generated by these fluxes is determined (through \eqref{effpot}) by the K\"ahler potential
\bee
 K = K_k + K_Q
\ee
and by the superpotential:
\bea
 W &=& \int_Y \left( \Omega_c \wedge H^f + F^f_6 + J_c\wedge F_4^f - \frac{1}{2} J_c \wedge J_c \wedge F_2^f - 
	\frac{m}{6} J_c \wedge J_c \wedge J_c \right)
\eea
This superpotential depends, in general, on all the geometric moduli at tree-level. The system of equations governing supersymmetric vacua is:
\bee
  D_{t^a}W = D_{n^k}W = D_{T_\lambda}W = 0
\ee
In \cite{DeWolfe:2005uu} it was shown that under reasonably general assumptions, one can stabilise all geometric moduli in these constructions. The same considerations show that in the leading approximation, $h^{2,1}_+$ axions will remain unfixed. These solutions can moreover be brought into a regime where $g_s$ is arbitrary small and the volume is arbitrary large.

\subsection{Relation between M-theory and Type IIA Vacua}\label{MthIIA}

In \cite{Kachru:2001je}, it was argued that for a special class of $G_2$ compactification manifolds $X$, Type IIA orientifolds appear at special loci in their moduli space. More precisely, these $G_2$ manifolds have to be such that they admit the form $X=(Y\times \Ss^1)/\hat{\sigma}$, where $Y$ is a CY 3-fold and $\hat{\sigma}=(\sigma,-1)$ is an involution which inverts the coordinates of the circle $\Ss^1$ and acts as an antiholomorphic isometric involution on $Y$. $\sigma$ and $\hat{\sigma}$ can have non-trivial fixed points making $X$ a singular $G_2$ manifold. 

The $G_2$ embedding of Type IIA orientifolds can be found in \cite{Grimm:2004ua}.
The invariant 3-form $\Phi$ can be written in terms of the invariant forms defining the CY structure ($J$ and $\Omega$). The relation is given by:
\bee\label{MIIAPhi}
 \Phi=J\wedge \dd y^7 + 2 \re (\mathcal{C}\Omega)
\ee
where $y^7$ is a coordinate along $\Ss^1$ and the 1-form $\dd y^7$ is normalised such that $\int_{\Ss^1}\dd y^7 = 2\pi R$, with $R$ the dilaton independent radius of the internal circle.
Substituting it in the K\"ahler potential \eqref{MthKpot}, one gets:
\bee
 K = -\ln\left(\frac{1}{6}\int_Y J\wedge J\wedge J\right) - 2 \ln \left(2\int_Y \re(\mathcal{C}\Omega)\wedge \ast_6\re (\mathcal{C}\Omega)\right)
\ee
This is exactly the form of the Type IIA K\"ahler potential $K_k+K_Q$.

In order to find how the superpotential transforms one needs the expressions for the potential $C$ and the field strength $G$:
\bea\label{MIIACG}
 C = B\wedge \dd y^7 +  C_3  &&  G = \dd C + H^f\wedge \dd y^7 +  F^f_4 
\eea
Substituting \eqref{MIIAPhi} and \eqref{MIIACG} into \eqref{MthSuperpot},\eqref{Mthsuppot}, one finds:
\bee
 W = \int_Y \left(J_c\wedge F^f_4 + \Omega_c\wedge H^f\right) + c_1 + i c_2
\ee
This is different from what was found in \cite{DeWolfe:2005uu}. At first, the fluxes $m$ and $F_2^f$ are zero; actually they should arise once manifolds of $G_2$ structure (instead of $G_2$ holonomy) are considered. Secondly, there is a constant term $c_1+ic_2$. In M-theory it comes from non-Abelian gauge fields living on the singularities. These singularities are dual to D6-branes, so this term must arise from non-Abelian background values of gauge fields living on D6-branes, as argued in \cite{Acharya:2002kv}.

The relation between the moduli is very simple. The M-theory moduli come from the expansion of $C+i\Phi$, while the Type IIA moduli from the expansions of $J_c$ and $\Omega_c$. Using the relations \eqref{MIIAPhi} \eqref{MIIACG} one obtains:
\bee
 z_i \phi_i = C+i\Phi = J_c \wedge dy^7 + \Omega_c = t_a \omega_a \wedge dy^7 + 2n^k\alpha_k + i T_\lambda\beta^\lambda
\ee
This gives the identification $z^a = t_a$ and $z^K = (n^k,T_\lambda)$ \cite{Grimm:2004ua}. Analogous identifications can be done for flux parameters.

One important consequence is that, when it is written in terms of flux parameters and moduli, the potential is formally the same for M-theory and Type IIA.

\section{Type IIA on Orientifolds}

In this section we will consider compactification of Type IIA on the orbifold limit of a CY. We will turn on $H$ and $F_4$ fluxes and the constant $c = c_1 + i c_2$. We are able to compute the explicit potential. Minimising it we will find all the supersymmetric and non-supersymmetric vacua.

\subsection{Toroidal Orbifolds}\label{TorOrb}

A six dimensional toroidal orbifold is the space resulting from the modding of the torus $T^6$ with a discrete isometry group $\Gamma$: $Y=T^6/\Gamma$. When this group does not act freely, we have a proper orbifold, {\it i.e.} a singular space.
We consider only orbifolds with Abelian point group. The action of the generator $\theta$ of $\Zbb_N$ on the torus $T^6$ is given by the 3-vector $(u_1,u_2,u_3)$ (with $0\leq u_i <1$):
\bea
 \theta : (w^1,w^2,w^3) &\mapsto& (e^{2\pi i u_1} w^1,e^{2\pi i u_2}w^2,e^{2\pi i u_3}w^3)
\eea
where $w^i$ are complex coordinates on $T^6$.
To obtain $\NN=2$ supersymmetry in four dimensions (before orientifolding), the group $\Gamma$ must be a subgroup of $SU(3)$, to furnish $SU(3)$ holonomy and to give a singular limit of a CY. This requires $\pm u_1 \pm u_2 \pm u_3 =0$ (see Ch. 8,9 in \cite{Blumenhagen:2006ci}). Implying also that $\Gamma$ must act crystallographically on the lattice specified by $T^6$, leads to relative few choices for $\Gamma$, {\it i.e.} $\Gamma=\Zbb_N$ with $N=3,4,6,7,8,12$ or $\Gamma=\Zbb_M\times\Zbb_N$ with $M$ multiple of $N$ and $N=2,3,4$. $\Zbb_6$, $\Zbb_8$ and $\Zbb_{12}$ have two different embedding in $SO(6)$.

The modding by the group $\Gamma$ cuts some of the moduli of the torus. In a CY compactification, we have $h^{1,1}$ K\"ahler moduli and $h^{2,1}$ complex structure moduli. A six dimensional torus has nine (1,1) harmonic forms and nine (2,1) harmonic forms. The orbifold group $\Gamma$ projects out some of the harmonic forms. The so called {\it untwisted moduli} of the orbifold compactification are related to these surviving forms. The twist elements $\theta,...,\theta^{N-1}$ produce conical singularities at the fixed points. In a small neighbourhood around them, the space is locally $\Cbb^3/\Gamma$ (isolated singularity) or $\Cbb^2/\Gamma^{(2)}\times \Cbb$ (non-isolated singularity). Roughly speaking, the singularities are resolved by substituting them with some new cycles. This introduces other geometric moduli that are called {\it twisted moduli}.  

In what follows, we will consider the orbifold limit of the resolved CY, which means that we will take the twisted moduli small compared to the untwisted one, and we will neglect them. At the end of calculation, we will justify this procedure, showing that the twisted moduli can be stabilised at higher energy than the untwisted ones, just by tuning fluxes on the cycles resolving the singularities. 

Let us then concentrate on the untwisted moduli. The (1,1) form on $T^6$ are the nine forms $dw^i\wedge d\bar{w}^{\bar{j}}$. If all the $u_i$'s are different from each other, only the three forms $dw^i\wedge d\bar{w}^{\bar{i}}$ survive the orbifold projection. If two of the $u_i$'s are equal to each other, say $u_1=u_2$, also the two forms $dw^1\wedge d\bar{w}^{\bar{2}}$ and $dw^2\wedge d\bar{w}^{\bar{1}}$ survive, giving $h^{1,1}_{untw}=5$. The case of $\Zbb_3$ is particular, because it has $(u_1,u_2,u_3) = (1/3,1/3,-2/3)$ and so the phases $e^{2\pi i u_i}$ are all equal to each others, giving $h^{1,1}_{untw}=9$. 
With analogous arguments, one can show that the only possible values for $h^{2,1}_{untw}$ are $0$ or $1$ for all the orbifold considered above, except for $\Zbb_2\times \Zbb_2$ that has $h^{2,1}_{untw}=3$.

We now introduce an orientifold 06-plane. It fills the four spacetime dimensions and wraps a supersymmetric 3-cycle. It gives $\NN=1$ supersymmetry in four dimensions. The 3-cycle is the fixed point set of the antiholomorphic involution $\sigma$. The orientifold projection cuts part of the moduli that survived the orbifold one. 

Under $\sigma$, the forms $dw^i\wedge d\bar{w}^{\bar{i}}$ are odd and so they are projected in. Also one linear combination of $dw^1\wedge d\bar{w}^{\bar{2}}$ and $dw^2\wedge d\bar{w}^{\bar{1}}$ is odd, while the orthogonal one is even. So the possible values for $h^{1,1}_{(-)untw}$ are $4$ or $3$ (we are neglecting the $\Zbb_3$ case). As regard the complex structure moduli, we have seen above that only one real component of each complex modulus survives.
 
Thus, for all the orientifolds of the orbifold models listed above (apart for $\Zbb_3$ and $\Zbb_2\times\Zbb_2$), the number of untwisted K\"ahler moduli is $3$ or $4$, while the number of complex structure moduli is $0$ or $1$.
We will consider first the case of $h^{1,1}_{(-)untw}=3$ and then we will treat the case $h^{1,1}_{(-)untw}=4$ in more details.

\subsubsection{$h^{1,1}_{(-)untw}=3$: dual to M-theory vacua}

In the case when $h^{1,1}_{(-)untw}=3$, the K\"ahler form is expanded as
\bee
 J = \sum_{i=1}^3 v_i\, \omega_i
\ee
where $\omega_i\propto dw^i\wedge d\bar{w}^{\bar{i}}$. The K\"ahler potential for the K\"ahler moduli is:
\bee
 K_k = - \ln \left( \kappa v_1 v_2 v_3 \right)
\ee
The 4-form $F_4^f$ is even under $\sigma$ and is expanded on the dual basis $\tilde{\omega}_i$ of $\omega_i$: $F_4^f=N_i \tilde{\omega}_i$.

Let us see what happens to the complex structure moduli sector.
\begin{itemize}
\item When $h^{2,1}_{untw}=0$, the third cohomology group has dimension equal to 2. Its basis is made up of two elements: $\alpha_0$, that is even, and $\beta_0$, that is odd. The expansion of $\Omega$ on this basis is then given by:
\bee
 \Omega = \frac{1}{\sqrt{2}} \alpha_0 + i \frac{1}{\sqrt{2}} \beta^0
\ee
where we have imposed the normalisation $i \int_Y \Omega\wedge \bar{\Omega} =1$. From here and \eqref{OmegaC3exp} we can identify: $Z^0=i\FF_0=\frac{1}{\sqrt{2}}$. Actually there are no complex structure moduli, and the dilaton is the only modulus appearing in $K_Q$, that takes the form:
\bee\label{KQmodel}
 K_Q = -2 \ln (\frac{1}{2} \ell^2)
\ee
where $\ell \equiv 2 \im n^0$ (see \eqref{Nkdef}); in this particular case $\im n^0 = \frac{e^{-D}}{\sqrt{2}}$.
Since $h^{2,1}_{untw}=0$, the 3-form $H^f$ (that is odd under $\sigma$) is proportional to $\beta^0$: $H= p \beta^0$.

In summary, when $h^{1,1}_{(-)untw}=3$ and $h^{2,1}_{untw}=0$, the K\"ahler potential and the superpotential are given by:
\bea\label{Kh113}
 K &=& K_k+K_Q = - \ln \left(\frac{1}{4}\kappa \, v_1 v_2 v_3 \ell^4\right) = - 3 \ln \left(\frac{1}{4}\kappa \, v_1 v_2 v_3 \ell^4\right)^{1/3}\\
 W &=& W_1+iW_2 = (\xi N_\ell + N_i b_i + c_1) + i (\ell N_\ell + N_i v_i + c_2)
\eea
With the dictionary given in Section \ref{MthIIA} , we see that these expressions are written, in M-theory language, as:
\bea
 &&K = - 3 \ln (V_X)  \,\,\, ;\,\,\, V_X =  s_1^{1/3} s_2^{1/3} s_3^{1/3} s_4^{4/3} \\
 &&W = (\sum_{j=1}^{4} N_j z_j) + c_1 + i  c_2 
\eea
(we have rescaled the $v_i$, such that $s_i= (\kappa/4)^{1/3}v_i,s_4=\ell$ ). 
The potential coming from these $K$ and $W$ is the same studied in \cite{Acharya:2005ez}, with $b^3=4$, $a_1=a_2=a_3=1/3$ and $a_4=4/3$.

\item When $h^{2,1}_{untw}=1$, the third cohomology group has dimension equal to 4. In this case, under the orbifold and the orientifold projections only the real component $U$ of one complex structure modulus survives. In particular for these cases we can write $w^1=x^1+iy^1$, $w^2=x^2+iy^2$ and $w^3=x^3+i U y^3$. Knowing that $\Omega\propto dw^1\wedge dw^2 \wedge dw^3$ and imposing the condition $i\int_Y \Omega\wedge \bar{\Omega} =1$, we can expand the holomorphic 3-form on a basis of real harmonic 3-forms ($\alpha_0$ and $\alpha_1$ even, $\beta^0$ and $\beta^1$ odd) as:
\bee
 \Omega = \frac{1}{\sqrt{U}} \alpha_0 + 2\sqrt{U}\alpha_1 + i \frac{\sqrt{U}}{4} \beta^0+ i\frac{1}{8\sqrt{U}}\beta^1
\ee
With the same procedure as above, one finds:
\bee\label{KpotQ2}
 K_Q = -2 \ln (\frac{1}{8} \ell^0 \ell^1)
\ee
where $\ell^0 \equiv 2 \im (n^0) = \frac{e^{-D}}{\sqrt{U}}$ and $\ell^1 \equiv 2 \im (n^1) = 2 e^{-D} \sqrt{U}$.

Following the steps described at the previous point, one can find that the K\"ahler potential and the superpotential of Type IIA can be written, in M-theory language as:
\bea
 &&K = - 3 \ln (V_X)  \,\,\, ;\,\,\, V_X =  s_1^{1/3} s_2^{1/3} s_3^{1/3} s_4^{2/3} s_5^{2/3} \\
 &&W = (\sum_{j=1}^5 N_j z_j) + c_1 + i  c_2 
\eea
The potential is the one given in \cite{Acharya:2005ez}, with $b^3=5$, $a_1=a_2=a_3=1/3$ and $a_4=a_5=2/3$.
\end{itemize}

\subsubsection{$h^{1,1}_{(-)untw}=4$: new vacua}

We have seen that when $u_1=u_2$, the basis of harmonic (1,1)-forms is made up of one more elements, with respect to the previous case, {\it i.e.} $\omega_4 \propto (dw^1\wedge d\bar{w}^{\bar{2}}-dw^2\wedge d\bar{w}^{\bar{1}})$. The K\"ahler form is expanded on this basis:
\bee
 J = \sum_{i=1}^4 v_i\, \omega_i
\ee
The K\"ahler potential for the K\"ahler moduli is then:
\bee\label{Kpotnew}
 K_k = - \ln \left( \kappa v_1 v_2 v_3 - \frac{1}{2}\kappa v_3 v_4^2 \right)
\ee
The 4-form $F_4^f$ is expanded as before on the four basis elements $\tilde{\omega}_i$: $F_4^f=N_i\tilde{\omega}_i$. Its contribution to the superpotential is again given by $\sum_{i=1}^4N_i t_i$.

The discussion of the complex structure sector remains the same as the case of $h^{1,1}_{(-)untw}=3$.
The superpotential is given by the formulae above, but with one more term given by $N_4 v_4$.

The total K\"ahler potential is the sum of $K_Q$ as given in the previous subsection and the expression \eqref{Kpotnew} for $K_k$. If we translate it in M-theory language, we see that the resulting K\"ahler potential cannot be written in the form $-\ln V_X$ with $V_X$ given in \eqref{VxMth}. So this is the case that still needs to be considered in order to complete the study of the stabilisation of the untwisted moduli in orientifold of Type IIA orbifold compactification with all possible $\Zbb_N$ and $\Zbb_N\times \Zbb_M$ orbifold groups. In this work we will study the potential in the case $h^{2,1}_{untw}=0$ and we will argue that the case $h^{2,1}_{untw}=1$ is qualitatively similar.

This study is important not only for Type IIA, but also for M-theory flux compactifications. This is because the analysis of the potential is formally the same in the M-theory dual, and this dual is not included in the ensemble studied in \cite{Acharya:2005ez}. It is interesting to see if there are some modification of those results with a different K\"ahler potential.


\section{Type IIA Orientifolds  with $h^{1,1}_{(-)untw}=4$ and  $h^{2,1}_{untw}=0$\\ and M-theory duals}

\subsection{Moduli Stabilisation}\label{modu_stab}

In this section we will study the moduli potential for the ensemble of flux vacua relative to the orbifold models with $h^{1,1}_{(-)untw}=4$ and  $h^{2,1}_{untw}=0$. They are described by the K\"ahler potential and the superpotential given in the precious Section.

The K\"ahler potential is the sum of\eqref{KQmodel} and \eqref{Kpotnew}:
\bee
 K = -\ln\left( \kappa v_1 v_2 v_3  -\kappa v_3 (v_4)^2/2\right)  - \ln\left( \frac1 4 \ell^4 \right)
\ee

We define $\{s^I\}\equiv\{v_i,\ell\}$. From the K\"ahler potential we can compute the metric on the moduli space $g_{I\bar{J}} = \frac{1}{4}K_{IJ}$. Even though it is not of diagonal form, as that obtained for the case $h^{1,1}_{(-)untw}=3$, we can diagonalise it; the eigenvalues are given by:
\bea
 &&\left\{ \frac 4 {\ell^2}\, ,\, \frac{1}{v_3^2} \, ,\, \frac{2\big(v_1^2+v_2^2+v_4^2-\sqrt{(v_1^2+v_2^2+v_4^2)^2-(v_4^2 -2 v_1 v_2)^2}\big)}{(v_4^2 -2 v_1 v_2)^2}, \right. \\&&\left. \frac{2\big(v_1^2+v_2^2+v_4^2+\sqrt{(v_1^2+v_2^2+v_4^2)^2-(v_4^2 -2 v_1 v_2)^2}\big)}{(v_4^2 -2 v_1 v_2)^2} \, ,\, \frac{2}{2 v_1 v_2 - v_4^2}\right\}
\eea
We want this metric to be positive definite. We note that all the eigenvalues are positive, except for the last; thus, the positiveness of the metric requires
\begin{equation}\label{constrKpos}
 	2 v_1 v_2 - v_4^2 > 0\,.
\end{equation}

\

The superpotential is given by:
\bea
 W &=& \sum_{i=1}^4 N_i t_i + N_5 L + c =  W_1 + i W_2 \nn\\
 	&=& ( \sum_{i=1}^4 N_i b_i +  N_5 \xi + c_1) + i ( \sum_{i=1}^4 N_i v_i + N_5 \ell + c_2)
\eea
where $t_i=b_i+iv_i$, $L=\xi+i\ell$ and $c=c_1+ic_2$.

The corresponding potential is obtained from the standard four dimensional supergravity expression \eqref{effpot}:
\begin{equation} \label{potential}
 V =  e^K \left( g^{I\bar{J}} F_I \bF_{\bar{J}} - 3|W|^2 \right),
\end{equation}
where we have defined $\{z^I\}\equiv\{t_i,L\}$ and $\{N_I\}\equiv\{N_i,N_5\}$ with
\begin{equation}
 F_I \equiv D_{z^I} W \equiv (\partial_{z^I} + \partial_{z^I} K) W = N_I
 + \frac{1}{2i} \partial_{s^I} K \, W.
\end{equation}
Defining further the derivative of $K$ as:
\begin{equation}
 K_I \equiv \partial_{s^I} K(s),  \qquad K_{IJ} \equiv \partial_{s^I} \partial_{s^J} K = 4 g_{I \bar J}
\end{equation}
we can write the potential as:
\begin{eqnarray}
 V &=& e^K \left( 4 K^{IJ} (\re F_I) (\re F_J) +  K^{IJ} K_I K_J W_1^2 - 3 W_1^2 -
 3 W_2^2 \right) \nonumber \\
 &=& e^K \left(4 K^{IJ} N_I N_J + 4 c_2 W_2  + 4 W_1^2 \right).  \label{Vsplit}
\end{eqnarray}
where we have used the expression for $W$ and the relations
\begin{equation} \label{Kid}
 \sum_i v_i \partial_iK_k  + \ell \partial_\ell K_Q = -7, \qquad \sum_j v_i \partial_i\partial_j K_k = - \partial_j K_k 
 	\qquad \ell \partial^2_\ell K_Q = - \partial_\ell K_Q
\end{equation}
Since $W_1$ is the only objects that depends on the axions and everything else in (\ref{Vsplit}) depends only on the geometric moduli, it is clear that any critical point of $V$ will fix
\begin{equation}
 \sum_i N_i b_i + N_5 \xi + c_1 = 0
\end{equation}
and therefore $W_1 =0 \Rightarrow \im F_I =0$. Apart from this, the axions are left undetermined, and they decouple from the geometric moduli. Non-perturbative effects will generate a potential for the axions: since they live on a compact space, they will be stabilised by this potential. From now on we will work on the slice of moduli space which does not include the axions, so we can write
\begin{equation} \label{Vreal}
 V = e^K \left( 4 K^{IJ} N_I N_J  + 4 c_2 W_2  \right)
\end{equation}

\

Supersymmetric solutions are determined by
\begin{equation}
F_I=D_IW=0 
\end{equation}
which in our case are given by
\bea
 N_i + \frac{1}{2} K_i W = 0  &&  N_5  + \frac{1}{2} K_\ell W = 0
\eea
Inserting the expressions for $K_i$, $K_\ell$ and $W$ and solving in the moduli, one gets one solution:
\bee
 \{v_1,v_2,v_3,v_4,\ell\} = \left\{\frac{2c_2 N_2}{5(N_4^2 - 2 N_1 N_2)}, \frac{2c_2 N_1}{5(N_4^2 - 2 N_1 N_2)}, -\frac{c_2}{5 N_3},
		\frac{2c_2 N_4}{5(N_4^2 - 2 N_1 N_2)}, -\frac{4 c_2}{5 N_5} \right\}
\ee

The other vacua are critical points of the potential, {\it i.e.} solutions of the equations:
\begin{equation}\label{dIV}
 	\der_I V = 0\,.
\end{equation}
Careful analysis of such equations shows that there are 11 distinct solutions with completely stabilised real moduli, including the aforementioned supersymmetric one (see the Appendix for details).

We will mainly focus on two solutions which are the most interesting to us: the supersymmetric one and a second non-supersymmetric $AdS$ solution which turns out to be the absolute minimum of the scalar potential and is given by
\bee
 \{v_1,v_2,v_3,v_4,\ell\} = \left\{\frac{4 c_2 N_2}{5(N_4^2 - 2 N_1 N_2)}, \frac{4 c_2 N_1}{5(N_4^2 - 2 N_1 N_2)}, -\frac{2 c_2}{5 N_3}, -\frac{4 c_2 N_4}{5 (N_4^2 - 2 N_1 N_2)}, -\frac{2 c_2}{5 N_5}\right\}
\ee

Physical consistency of the solutions and validity of the supergravity end effective field theory approximations constrains the parameters of the model ($c_2$ and the fluxes $N_i$). 
We will describe in detail such physical consistency conditions and what constraints they imply on the parameters of the two solutions mentioned above. In the Appendix we will briefly discuss the properties of the other nine solutions.

As noted in \eqref{constrKpos} positiveness of the kinetic matrix requires
\begin{equation}
 	2 v_1 v_2 - v_4^2 > 0
\end{equation}
which implies, on both the supersymmetric vacuum and the AdS absolute minimum:
\begin{equation}
 	2 N_1 N_2 - N_4^2 > 0\,.
\end{equation}
We now introduce the variable
\begin{equation}\label{mbound}
 	m = \frac{N_4^2}{2 N_1 N_2}
\end{equation}
in terms of which the previous inequality is expressed as
\begin{equation}
 	0<m<1\,.
\end{equation}

The second requirement is the positiveness of the compactification volume. In string frame and in string units, it is given by:
\begin{equation}
\vol = \kappa v_1 v_2 v_3  -\kappa v_3 v_4^2/2= \frac{\kappa v_3}{2} ( 2 v_1 v_2  - v_4^2)
\end{equation}
Taking into account the condition \eqref{constrKpos}, this implies $\kappa v_3 >0$. The corresponding constraint on our two solutions is
\begin{equation}
 	\kappa c_2 N_3 < 0
\end{equation}

A third condition comes from the dilaton modulus $\ell$. It is related to the four dimensional dilaton $D$ and the ten dimensional dilaton $\phi$ as
\begin{equation}
 	\ell = \e^{-D} = {\vol}^{1/2} \e^{-\phi}
\end{equation}
which has to be a positive quantity. For both solutions we have:
\begin{equation}
 	c_2 N_5 <0
\end{equation}

The value of the scalar potential on the two extremal points we are considering is given by
%
\begin{equation}
 	V=\begin{cases}
 	   	\frac{9375}{32}\frac{(m-1) N_3 N_4^2 N_5^4}{c_2^5 \kappa m}\qquad \mbox{susy}\\
		\frac{3125}{8}\frac{(m-1) N_3 N_4^2 N_5^4}{c_2^5 \kappa m}\qquad \mbox{min}
 	  \end{cases}
\end{equation}
Taking into account all the conditions we have derived up to now, this shows that the two solutions are both $AdS$. The case $N_4=0$ is excluded because it gives $v_4=0$. Once given all the physical solutions, which are reported in the Appendix, we can easily conclude, by direct comparison, that the second solution is indeed an absolute minimum of the potential.
%
%
%
%
 
We now come to the issue of stability of the solutions that we obtained. An $AdS$ critical point $S$ does not have to be a local minimum to be perturbatively stable. It is sufficient that the eigenvalues of the Hessian of $V$ are not too negative compared to the cosmological constant, and more precisely that the Breitenlohner-Freedom bound \cite{Breitenlohner:1982bm} is satisfied:
\begin{equation}
 \frac{\der}{\der \hat{s}^H}\frac{\der}{\der \hat{s}^K}   V|_{S}-\frac{3}{2}V|_{S} \delta_{HK} > 0
\end{equation}
%
The derivatives are taken with respect to the scalars with canonically normalised kinetic term. This requires to diagonalise the K\"ahler metric and rescale the moduli. Some simple linear algebra considerations\footnote{The details are given in the Appendix}, allows us to reduce the above condition to the simpler one
\begin{equation}\label{stabilAdS}
 	\der_I \der_J V|_{S} - \frac 3 4 V|_{S} K_{IJ}|_{S} > 0
\end{equation} 
where now the derivatives are done with respect to the original moduli. A necessary and sufficient condition for a finite size matrix to be positive definite is that the determinants of all the diagonal minors are positive. Calculation of these determinants is not hard and shows that both vacua are stable.

We briefly summarise here the results for the generic solutions to the vector equation \eqref{dIV}  which are described in greater detail in the Appendix. There are in total 11 physical extremal points of the potential. Nine of them are $AdS$ vacua but only eight satisfy the condition in \eqref{stabilAdS}, for generic values of $c_2, N_I$. The remaining two extremal points are $dS$ vacua. In order to be metastable they must be local minima of the potential: both of them are unstable, presenting tachyonic directions. This also happens for $dS$ vacua in models with $h^{1,1}_{(-)untw}=3$, as proven in \cite{Acharya:2005ez}. So in the class of models we are analysing, there are no stable $dS$ vacua.

%

\subsubsection*{Twisted Moduli}

We now come to a brief analysis of the contribution of the K\"ahler twisted moduli\footnote{The analysis for the twisted complex structure moduli is analogous and we will not report it here.}. If we take for example the $T^6/\mathbb Z_4$ orbifold \cite{Ihl:2006pp}, their introduction modifies $K$ and $W_2$ as
\begin{multline}
 	K_{tw}= -\ln\left( \kappa v_1 v_2 v_3  - \kappa v_3 (v_4)^2/2+\frac{v_3 \alpha}{2}\sum_{A=4}^{14} v_A^2 + \frac{\beta}{6}\sum_{B=15}^{26} v_B^3\right)  - \ln\left( \frac1 4 \ell^4 \right)=\\=
 K  -\ln\left[ 1 + \e^{K_k} \left( \frac{v_3 \alpha}{2}\sum_{A=4}^{14} v_A^2 + \frac{\beta}{6}\sum_{B=15}^{26} v_B^3\right)\right]
\end{multline}
\begin{gather}
	W_2^{tw} = W_2 + \sum_{A=4}^{14} N_A v_A + \sum_{B=15}^{26} N_B v_B
\end{gather}
In our analysis, we have considered that
\begin{equation}
 	K_{tw} \approx K\qquad W_2^{tw} \approx W_2
\end{equation}
This approximation is reliable as long as some constraints on the values of the fluxes $(N_A,N_B)$ associated with the resolved cycles are satisfied. Let us consider the corrections to the supersymmetric vacuum which is determined by the equation
\begin{equation}
 	\der_I W^{tw} + \frac 12 \der_I K^{tw} W ^{tw}= 0
\end{equation}
For $I=1,\ldots, 5$, if we neglect the contribution of the twisted moduli, we obtain the previous solution. Focusing on the equation for $I=A,B$ we can check the consistency of our approximation. The leading order expansion gives
\begin{equation}
 	\der_A W_2^{tw} + \frac 12 \der_A K W_2 \approx N_A - \frac 12\alpha \e^{K_k}v_3 v_A W_2
\end{equation}
We recall now that, neglecting the twisted moduli
\begin{equation}
 	v_I \sim \frac{c_2}{N} \qquad I=1,\ldots,5
\end{equation}
where $N$ is the typical scale of the fluxes $\{N_I\}_{1\leq I\leq 4}$. We thus conclude that
\begin{equation}
 	v_A \sim \frac{c_2}{N} \frac {N_A} N
\end{equation}
A similar analysis for $I=B$  shows
\begin{equation}
 	v_B \sim \frac{c_2}N \frac {N_B^{1/2}}{N^{1/2}}\,.
\end{equation}
It is thus enough to choose
\begin{equation}
 	N_{A,B} \ll N
\end{equation}
to have a consistent solution. 

For the non supersymmetric solution, one can perform a similar perturbative analysis considering the more complicated equations
$\der_I V$ and draw similar conclusions: the twisted moduli are fixed by fluxes on the resolving cycles, and if these fluxes are much smaller than the fluxes on the normal cycles, then the physical quantities are mainly determined by the untwisted moduli.

\subsection{Physical Consistency Checks}

In this section we study the validity of our supergravity and effective field theory treatment analysing the different parameters of the solutions. The quantities involved are the ten dimensional Planck scale $m_p$, the string
scale $m_s$, the compactification volume in Einstein frame $\Vol_E$, the compactification volume in String frame $\Vol_S$, the string coupling $g_s$ and the Kaluza-Klein scale $m_{KK}$ and the four dimensional Planck mass $M_4$. These are related to each other as follows:
\bea
 m_p^8 = \frac{M_4^2}{\Vol_E} &&  m_s^2 = g_s^{1/2}m_p^2  \nn\\
 \Vol_E = g_s^{-3/2} \Vol_S &&  m_{KK}^2 = \frac{1}{\Vol_E^{1/3}}
\eea

The other quantities involved are the cosmological constant $\Lambda$ and the gravitino mass $m_{3/2}$.

All these quantities depend on the moduli and must be evaluated on the vacua. In particular one can see
that those belonging to the first group depend only on $\vol = \kappa v_3(v_1v_2 - \frac12 v_4^2)$ and $\ell$. In particular, using the fact that $\Vol_S = \vol \cdot m_s^{-6}$ and that $g_s=\frac{\vol^{1/2}}{\ell}$, one gets:
\bea
 &&m_p^2 = \frac{\vol^{1/2}}{\ell^3} M_4^2 \qquad m_s^2 = \frac{\vol^{3/4}}{\ell^{7/2}} M_4^2 \qquad m_{KK}^2=\frac{\vol^{2/3}}{\ell^4} M_4^2 \nn\\
 &&\Vol_E = \frac{\ell^{12}}{\vol^2} M_4^{-6} \qquad \Vol_S = \frac{\ell^{21/2}}{\vol^{5/4}} M_4^{-6} 
\eea
All these quantities are written in units of the four dimensional Planck mass $M_4$.

We remember that the cosmological constant and the gravitino mass are given by:
\bea\label{CCandGravMass}
 \Lambda = V(s_I; N_I)\, M_4^4 && m_{3/2}^2 = \e^K(s_I) |W(s_I; N_I)|^2 \, M_4^2 
\eea
where $V$, $K$ and $W$ are dimensionless functions of the moduli and of the fluxes.

To control $\alpha'$ and $g_s$ corrections we must have 
\bea
 	\Vol_S\gg m_s^{-6} \: (\Longleftrightarrow \vol \gg 1)&,&  g_s \ll 1 
\eea
One can check that these two conditions imply $\Vol_E \gg m_p^{-6}$ and hence the validity of the supergravity approximation.

Moreover we require to have a meaningful four dimensional effective theory, decoupled from the KK modes. Hence we need that the Hubble scale $H$ defined by $H^2 = \frac{|\Lambda|}{M_4^2}$, is less than the KK scale $m_{KK}$. Putting $M_4=1$, this means:
\bee
 |\Lambda| \ll m_{KK}^2
\ee

In order to have decoupling of the moduli and the gravitino we must finally have
\begin{equation}
 	m_v, m_{3/2} \ll m_s, m_{KK}.
\end{equation}

We set logarithmic scales $\rho$ and $\lambda$ which fix the level of confidence of our approximation. The consistency conditions are thus:
\begin{equation}\label{consistency}
	\begin{split}
		\vol & > 10^{6\lambda}\\
	 	g_s&< 10^{-\rho}\\
		|\Lambda|/m_{KK}^2 &< 10^{ -2\lambda}\\
		m^2_{3/2} /m^2_s &< 10^{-2\lambda}\\
		m^2_{3/2} /m^2_{KK} &< 10^{-2\lambda}\\
	\end{split}
\end{equation}
We will discuss the conditions on $m_v$ later. 

\ 

We introduce new variables $\gamma,\mu$ and $n$ defined by
\begin{eqnarray}\label{newvariab}
 	 	\e^{\gamma} &=& |c_2|\nn\\
		\e^{\mu} &=& |N_3| \frac{(1-m)}{\kappa m}N_4^2\label{muDef} = 
			\kappa^{-1}|N_3|(2N_1N_2-N_4^2) \\
		\e^{n} &=& |N_5|\nn
\end{eqnarray}
Since generic values of the moduli for the two solutions under examination are proportional to $\frac{c_2}{N}$ with $N$ a generic flux, if $c_2$ is too small we have very small cycles and thus breakdown of $\alpha'$ expansion. For such reason we can assume $\gamma > 0$. 

One can study the consistency conditions \eqref{consistency} both for the supersymmetric vacuum and to the absolute minimum. Here we present the computations for the latter case. 

The physicalquantities which appear in \eqref{consistency} are expressed in terms of the variables \eqref{newvariab} as:
\bea\label{massesmugamma}
 \vol &=& \frac{16}{125}\e^{3\gamma-\mu} \nn\\
 g_s &=& \frac{2}{\sqrt{5}}\e^{n+\gamma/2-\mu/2} \nn\\
 |\Lambda| &=&  \frac{3125}{8}\e^{4n-5\gamma+\mu} \nn\\
 m_{3/2} &=&  \frac{28125}{64}\e^{4n-5\gamma+\mu} \nn\\
 m_{KK} &=&  \frac{25}{2^{4/3}}\e^{4n-2\gamma-2\mu/3} \nn\\
 m_s &=&  \frac{5^{5/4}}{\sqrt{2}}\e^{\frac14(14n-5\gamma-3\mu)} \nn
\eea

Substituting these expression in \eqref{consistency} and taking the logarithm, we reduce to a set of linear inequalities which involve only $\gamma,\mu,n,\lambda$:
\bea\label{LinIneq}
 \vol > 10^{6\lambda} &\rightarrow& \mu < 3\,\gamma -13.8\,\lambda  -2.1 \nn\\
 g_s< 10^{-\rho} &\rightarrow& \mu > \gamma +2\,n +4.6\,\rho  - 0,2 \nn\\
 |\Lambda|/m_{KK}^2 < 10^{ -2\lambda} &\rightarrow& \mu < 1.8\, \gamma -0.1\,\lambda  -2.2 \\
 m^2_{3/2} /m^2_s < 10^{-2\lambda} &\rightarrow& \mu < -0.3\, n +2.1\,\gamma -2.6\,\lambda -2.5 \nn\\
 m^2_{3/2} /m^2_{KK} < 10^{-2\lambda}&\rightarrow& \mu < 1.8\,\gamma -2.8\,\lambda  -2.3\nn
\eea
where we have approximated the numerical quantities, just to have an idea of the form of the inequalities.


\begin{figure}
\begin{center}
  \epsfig{file=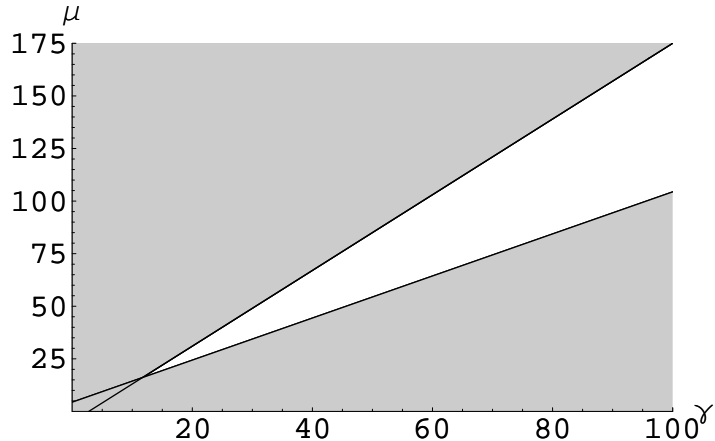,height=6cm,angle=0,trim=0 0 0 0}
  \caption{\small $\rho = 1$, $\lambda = 1$. The coloured region of the $(\phi,\mu)_{n=0}$ parameter space is excluded by the consistency conditions.}
  \label{fig}
\end{center}
\end{figure}

We will present a graphical analysis of such inequalities in the $\gamma-\mu$ and $n-\mu$ planes: one can draw a set of straight lines which represent the boundary of the region of validity of each inequality. Intersecting all such regions we will get the allowed subset of the space of parameters. First of all, we notice that 
taking larger and larger $n$ one restricts the region of validity of the inequalities. For this reason, we first study the problem for fixed $n=0$ (corresponding to $N_5=1$), {\it i.e.} its minimal value. 
In figure \ref{fig}, we show an example for fixed values of $\rho$ and $\lambda$. The two straight lines are relative to the conditions $g_s\ll 1$ and $|\Lambda|/m_{KK}^2 \ll 1$. Actually one can see that for the fixed values of $\lambda$, $\rho$ and $n$, requiring only $g_s\ll 1$ and $|\Lambda|/m_{KK}^2 \ll 1$ is enough to have all consistency conditions \eqref{consistency} satisfied.
The excluded region is coloured in grey. From this graph we see that we have acceptable vacua only for large values of $|c_2|=\e^\gamma$. In particular, for $\rho = 1$ and $\lambda = 1$  we have $|c_2|\gtrsim 10^{6}$.

\

\begin{figure}
\begin{center}
  \epsfig{file=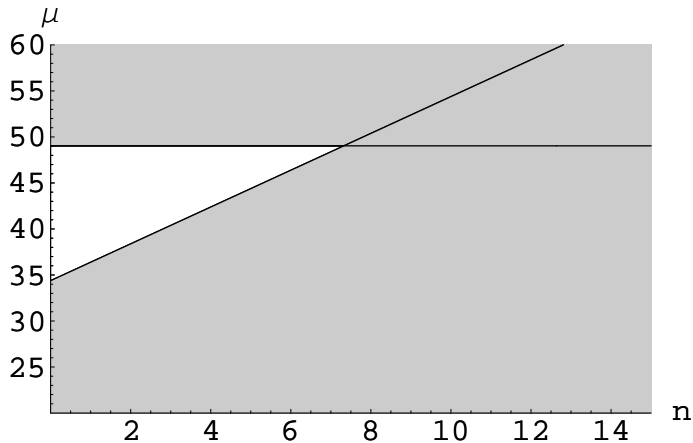,height=6cm,angle=0,trim=0 0 0 0}
  \caption{\small $\rho = 1$, $\lambda = 1$. The coloured region of the $(n,\mu)_{\gamma=30}$ parameter space is excluded by the consistency conditions.}
  \label{fig2}
\end{center}
\end{figure}

Then we fix one value of $\gamma$ and find the region of validity in the $n-\mu$ space. For $\rho = 1$, $\lambda = 1$ and $\gamma=30$ ($c_2\sim 10^{13}$), the allowed region is drawn in figure \eqref{fig2}.
From it we note that only solutions corresponding to extremely small values of $N_5$ are acceptable. 

\

So far, we have not considered the condition that the masses of the moduli should be smaller than the KK scale $m_{KK}$. Since we have neglected the KK modes, this is necessary for the consistency of our analysis. Moreover, we have stayed within the Ricci flatness approximation,  assuming that the fluxes do not modify sensitively the geometry of the 6D manifold, in such a way that we can still consider it a Calabi-Yau.  This approximation is justified as long as the fluxes give the otherwise massless scalars a mass that is small compared to the KK masses. In this way the light spectrum remains unaltered after the introduction of fluxes, apart for the masses acquired by the moduli. 

We compute the masses of the moduli in the absolute minimum. To do this computation, we assume that the uplift to a Minkowski vacuum (or dS with a tiny cosmological constant) has been done. 
In this case the mass matrix is given by $\hat{\der}_I\hat{\der}_JV$, with derivative done with respect to the canonically normalised scalars (see Appendix). The eigenvalues of this matrix are all of the same order of magnitude:
\bee
 m_v^2 = \e^{4n-5\gamma +\mu+7}
\ee
$m_v$ depends on fluxes and $c_2$, through the quantities $\mu$, $n$ and $\gamma$. Imposing the condition $m_v/m_{KK} < 10^{-\lambda}$, one gets a further linear relation:
\bee
 \mu < 1.8\,\gamma -2.8\,\lambda -4.2
\ee
One can draw this straight line in the graphs \ref{fig} and \ref{fig2}, and see that the allowed region stays in the region selected by \eqref{LinIneq}, giving no more constraints.   

%

\

From these computations we have obtained the region of the parameters space corresponding to consistent approximations. This region is populated densely (but discretely) by vacua, as one can see looking at the expression \eqref{muDef}.
More precisely, once one fixes $c_2$ and $n$, then there is a countable infinity of vacua that realize $\mu$ between the two bounds. This happens because the combination $(2N_1N_2-N_4^2)$ can remain relatively small even if the fluxes $N_1,N_2$ and $N_4$ become arbitrarily large.
If there were no bound on these fluxes, then there would exist an infinity of vacua with moduli stabilised and with a valid four dimensional description. This infinite number of vacua would be an important difference with respect to vacua coming from $h^{1,1}_{(-)untw}=3$ orbifold models. In that case, the consistency of $\alpha'$ expansion requires the moduli $v_i$ to be large. Since in that solution $v_i \sim c_2/N_i$, this implies an upper bound on $N_i$. 
This is true in the models studied here only for the $v_3$ modulus (we are referring to the two vacua on which we focused). This gives a bound on $N_3$ ($N_3\lesssim c_2$). On the other hand the fluxes $N_1,N_2$ and $N_4$ can be taken arbitrary large values, the only condition being that the combination $N_4^2-2N_1N_2$ remains smaller than $c_2N_1$, $c_2N_2$ and $c_2N_4$.

Luckily, there is one further consistency requirement that we have not considered yet. Supersymmetrisation of higher derivative curvature terms gives rise to terms proportional to powers of $|F_4|^2$.  Even keeping the combination $N_4^2-2N_1 N_2$ small, large volumes of the single $N_i$ will lead to large values $|F_4|^2$. For example, from the expression of the metric for the K\"ahler moduli we can easily derive what is contribution of a single power of $|F_4|^2$
\begin{equation}
 	\int F_4\wedge\star F_4 \propto 2N_1^4+2N_2^4+2N_3^4 + N_4^2 \left( N_4^2 + 4 N_1^2 +4 N_2^2 +4 N_1 N_2\right)
\end{equation}
In general we can expect that large values on $N_i$ will give rise to large fluxes; higher derivatives terms in the effective action are multiplied by powers of
\begin{equation}
 	\ell_s^2  \propto \left( 2 N_1 N_2-N_4^2\right)^{3/2}
\end{equation}
This says that the fluxes $N_1,N_2$ (and $N_4$) cannot be arbitrarily large.  With this condition, the number of reliable vacua becomes finite and a statistical analysis is possible. 

\

In Type IIB the bound on fluxes was given by the tadpole cancellation conditions on the D3-brane charge, that involved the 3-form fluxes. Here the tadpole cancellation conditions do not imply bound on fluxes. This is a difference also with respect to the Type IIA ensemble of vacua studied in \cite{DeWolfe:2005uu}, where the D6-charge cancellation implied constraint on the $H_3$ fluxes. Here this does not happen because $m=0$ and therefore $H_3$ does not enter in the D6 tadpole cancellation condition. However we have a bound on the $H_3$ flux, coming from requiring $\ell\gtrsim 1$: $N_5\lesssim c_2$. 

\subsubsection*{Consistency Checks in M-theory dual Vacua}

We have seen that the Type IIA vacua described above have M-theory dual vacua. 

We now consider the M-theory vacua dual to the Type IIA ones studied above. We have seen that the form of the potential is the same in two cases, but that the physical meaning of the quantities involved is different. We can simply translate the Type IIA solutions in M-theory language\footnote{
In finding the minima of the potential in Type IIA we imposed some conditions on the validity of the solutions, like the positiveness of the moduli space metric. The conditions coming from M-theory give the same constraints on the solutions.} 
and then see if these vacua are consistent with the supergravity approximation of M-theory and with neglecting the KK modes.

The volume of compactification is given in M-theory by 
\bee
 \Vol X = V_X m_P^{-7}
\ee
where $m_P$ is the eleven dimensional Planck scale. $V_X$ is given by \eqref{MthKpot}. Substituting the expansion of the 3-form $\Phi$ and knowing that $s_i=\kappa^{1/3}v_i$ for $i=1,2,3$ and $s_4=\ell$, we obtain: 
\bee
 V_X = (\vol \ell^4)^{1/3}
\ee

Now, we express everything in units of the four dimensional Planck mass $M_4$. From the relations $m_P^9 = \frac{1}{\mbox{VolX}} M_4^2$ and $m_{KK}=\frac{1}{\mbox{VolX}^{1/7}}$, one gets
\bee
 \Vol X = V_X^{9/2} M_4^{-7} \qquad m_P^2 = \frac{1}{V_X} M_4^2 \qquad m_KK^2 = \frac{1}{V_X^{9/7}} M_4^2
\ee
We see that they are all function of $V_X$. The cosmological constant and the gravitino mass are still given by \eqref{CCandGravMass}.

We introduce the variables $\mu,n$ and $\gamma$ \eqref{newvariab}. In terms of them 
\bee
 V_X = \frac{4}{5^{7/3}}\e^{\frac73 \gamma}\e^{-\frac13(4n+\mu)}
\ee
while the cosmological constant and the gravitino mass are given by:
\bea
 |\Lambda| = \frac{3125}{8}\e^{-5\gamma}\e^{4n+\mu} && m_{3/2} = \frac{28125}{64}\e^{-5\gamma}\e^{4n+\mu} 
\eea
We see that for fixed $c_2$ ($=\e^{\gamma}$) all these quantities depend on the fluxes through the combination 
\bea
 \e^\nu \equiv \e^{4n+\mu} = N_5^4 |N_3|(2N_1N_2-N_4^2)
\eea

\

\begin{figure}
\begin{center}
  \epsfig{file=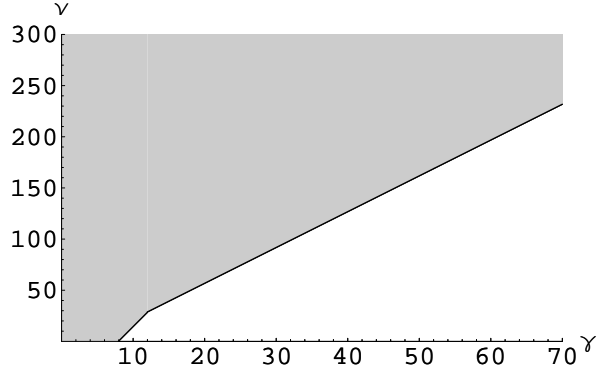,height=5cm,angle=0,trim=0 0 0 0}
  \caption{\small $\lambda = 1$. The coloured region of the $(\gamma-\nu)$ parameter space is excluded by the consistency conditions.}
  \label{fig4}
\end{center}
\end{figure}

The consistency conditions are given by the reliability of the supergravity approximation, by the decoupling of KK modes and by asking the gravitino mass to be small
with respect to the KK scale:
\bee
 \Vol X \gg m_P^{-7} \qquad \frac{|\Lambda|}{M_{4}^2} \ll m_{KK}^2 \qquad m_{3/2} \ll m_{KK} 
\,.
\ee
Substituting into these relations the expressions in terms of $\gamma$ and $\nu$, we get conditions on $\gamma$ and $\nu$. Following an analogous procedure as in the previous Section, we obtain the allowed region in the $\gamma-\nu$ plane (see figure \ref{fig4}). The requirement on small moduli mass with respect to $m_{KK}$ is automatically satisfied in this region.

Again there is a lower bound on $c_2$, that in this case is $c_2\sim 2000$. We note that there is a range of $c_2$ values for which there are M-theory vacua, but not Type II vacua consistent with the approximations. 

As for Type IIA the region of validity is densely populated by vacua. This happens because we did not impose a bound on $N_1, N_2, N_4$, but only on the combination $(N_4^2-2N_1 N_2)$. In type IIA we obtained a bound on the values of $N_1,N_2,N_4$ from the analysis of higher derivative terms in the action. On physical grounds, these in general become relevant whenever the energy of the fluxes is large enough. Similar considerations should be applied to the M-theory case excluding thus large values of flux numbers.

%
%
%

\subsection{Gravitino Mass and Moduli Masses}

The masses of the gravitino and of the moduli and the cosmological constant take the same expression in both Type IIA vacua and M-theory ones. They all are proportional to 
\bee\label{mgravmod}
 M = c_2^{-5/2} \e^{2n+\mu/2} = c_2^{-5/2} \e^{\nu/2}
\ee
and they are of the order $M \cdot 10^3$. 
\begin{figure}
\begin{center}
  \epsfig{file=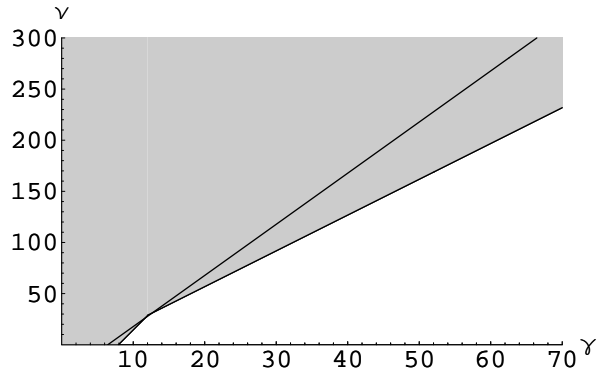,height=5cm,angle=0,trim=0 0 0 0}
  \caption{\small $\lambda = 1$. The straight line correspond to $M=10^{-7}$.}
  \label{fig3}
\end{center}
\end{figure}
One can compute what is the maximal value of $M$ in both the ensembles of vacua. To do this 
we take the logarithm of this expression, finding one straight line in the $\gamma-n-\mu$ space for each value of $M$. 
In this way one can see what are the possible values that can be realized in these vacua. Let us be more specific. Consider the M-theory case, with $\lambda=1$. Consider the figure \ref{fig3}. We have drawn the straight line corresponding to $M^{max}=10^{-7}$. Above this line we have points that realize bigger values of $M$, while below this line $M$ is smaller than $M^{max}$. So  the maximal value of the gravitino and moduli masses is given by 
\bee
m^{max}_{Mth} = 10^3 M^{max} M_4 \sim 10^{15} GeV
\ee

An analogous procedure can be applied to Type IIA vacua, giving 
\bee
m^{max}_{IIA} \sim 10^{12} GeV
\ee

As one can see from the analytic expression \eqref{mgravmod}, increasing the value of $c_2$ we can get smaller values for $M$. In particular for large $c_2$ there are vacua realizing small gravitino mass. 

We can be more explicit. Looking at the expressions \eqref{mgravmod}, we see that once fixed $c_2$, its dependence on the fluxes is only through $\e^\nu$. Knowing the distributions of this quantity over the set of vacua, one can find the distribution of for example $m_{3/2}$ or $\Lambda$. In particular one can count how many vacua there are corresponding to a given value of $\nu$. As we have seen previously, we get a finite answer, because the fluxes are bounded from above.
The distribution of $\e^\nu$ is really different with respect to the $h^{1,1}=3$ case, where it is substituted by a product of integers $\e^\nu=N_1N_2N_3N_4^4$. In that case, small values of $\e^\nu$ are clearly suppressed with respect to large values, especially when the fluxes can take very large values. In the present case, $\e^\nu$ is the product of two integers times the difference of two integers. This makes the distribution of $\e^\nu$ peaked around small values and makes more simple to get, for example, small gravitino mass 
and so intermediate supersymmetry breaking scale $M_s^2 \sim m_{3/2}\, M_4$.
This considerations are valid both in Type IIA and in M-theory. In the last case quantities like $M_s$ and $\Lambda$ depend on the volume $V_X$ of the seven dimensional manifold \cite{Acharya:2005ez}. When $V_X$ is given by \eqref{Kh113} the distributions were studied in \cite{Acharya:2005ez}, where it was found that small values for $M_s$ are suppressed. If instead of \eqref{Kh113}, one uses 
\bee\label{newMthVx}
V_X = s_3^{1/3}(s_1s_2-\frac{1}{2}s_4^2)^{1/3}s_5^{4/3}
\ee
then the distribution will be peaked around smaller values of $M_s$. This is an important difference that shows that changing the K\"ahler potential can sensibly change the distributions of physical quantities.

\subsection{The other new Case: $h^{1,1}_{(-)untw}=4$ and  $h^{2,1}_{untw}=1$}\label{modu_stab2}

To complete the study of orbifold models listed in Section 3 (apart for $\Zbb_3$ and $\Zbb_2\times\Zbb_2$), one should analyse the case $h^{2,1}_{untw}=1$. We do not expect qualitatively different results, being the form of the K\"ahler potential similar to the one we have already studied.

We can easily compute the supersymmetric vacuum for this case. The K\"ahler potential is given by $K=K_k+K_Q$ with $K_k$ given in \eqref{Kpotnew} and $K_Q$ in \eqref{KpotQ2}. The superpotential is 
\bea
 W &=& ( \sum_{i=1}^4 N_i b_i +  N_5 \xi_1 + N_6 \xi_2 + c_1) + i ( \sum_{i=1}^4 N_i v_i + N_5 \ell_1 +N_6 \ell_2 + c_2)
\eea
The solution of $D_IW=0$ is 
\bee
 \{v_1,v_2,v_3,v_4,\ell_1,\ell_2\} = \left\{\frac{2c_2 N_2}{N_4^2 - 2 N_1 N_2}, \frac{2c_2 N_1}{N_4^2 - 2 N_1 N_2}, -\frac{c_2}{5 N_3},
		\frac{2c_2 N_4}{N_4^2 - 2 N_1 N_2}, -\frac{2 c_2}{5 N_5}, -\frac{2 c_2}{5 N_6} \right\}
\ee
This is a metastable vacuum. 
The K\"ahler moduli, as functions of the fluxes and $c_2$ are stabilised at the same values as in the previous case.  It would be interesting to compute the full set of solutions of $\partial_IV=0$ and see what is the absolute minimum of the potential and if it presents the same features as we found for $h^{2,1}_{untw}=0$.

\section{M-theory without Fluxes}

In the previous section we have described Type IIA vacua, dual to flux M-theory vacua with a nonzero complex Chern-Simons invariant. Such kind of vacua have been studied in \cite{Acharya:2005ez} using the K\"ahler potential given by \eqref{VxMth}
\bea\notag
 V_X=\prod_{i=1}^{b^3}(s_i)^{a_i} &{\rm with }& \sum_{i=1}^{b^3}a_i = \frac{7}{3}\,.
\eea
That choice was justified by the fact that \eqref{VxMth} satisfies the necessary condition to be the volume of a $G_2$ holonomy manifold. 
In this work we have studied the same kind of vacua in type IIA, with a different choice of K\"ahler potential, in the case of 5 moduli. In M-theory language, the new K\"ahler potential satisfies the conditions required by $G_2$ holonomy. 

In \cite{Acharya:2007rc,Acharya:2006ia} the authors studied a further ensemble of M-theory vacua. They considered zero flux background, and studied the potential generated by a non-perturbative superpotential and the K\"ahler potential given by \eqref{VxMth}

The nonperturbative superpotential arises from strong gauge dynamics of a hidden sector residing on a set of three dimensional submanifolds $Q_i$. In M-theory, these submanifolds generically do not intersect each other and do no intersect the standard model one; for this reason, supersymmetry breaking is gravity mediated in these models. In \cite{Acharya:2007rc,Acharya:2006ia} it was showed that the resulting potential generically stabilises all the moduli and all supersymmetric and non-supersymmetric minima were found. The superpotential appearing there is:
\bee\label{BobKanSuperpot}
 W = A_1 \e^{ib_1\sum_I N_I^{(1)} z_I} +A_2 \e^{ib_2\sum_I N_I^{(2)} z_I}
\ee
where $b_k=2\pi/c_k$ with $c_k$ the dual Coxeter numbers of the hidden sector gauge groups, and $A_k$ are numerical constant. The linear combinations $\sum_I N_I^{(k)} z_I$ ($z_I=t_I+is_I$) are the gauge coupling functions of the hidden sector gauge groups.

In the case $N_I^{(1)}=N_I^{(2)}$ the supersymmetric vacuum is given \cite{Acharya:2007rc,Acharya:2006ia} by
\bea\label{BobKanSusyVac}
 s_I = \frac{a_I\nu}{N_I} &\mbox{ where }& \nu = -\frac{3(\alpha-1)}{2(\alpha b_1-b_2)}\\
 &\mbox{ with }& \frac{A_2}{A_1}\frac 1\alpha \e^{\frac72 (b_1-b_2)\frac{\alpha-1}{\alpha b_1-b_2}}
\eea

\

It would be interesting to study the potential that arises using the non-perturbative superpotential \eqref{BobKanSuperpot} and the K\"ahler potential given by \eqref{newMthVx} and see if one finds the same qualitative properties found in \cite{Acharya:2007rc,Acharya:2006ia}. Here we simply compute the supersymmetric vacuum. The supersymmetry equations $D_iW$ are given by:
\bea
 (b_1N_I^{(1)}-\frac12 K_I)A_1+(b_2N_I^{(2)}-\frac12 K_I)A_2 \e^{(b_1 \vec{N}^{(1)}-b_2 \vec{N}^{(2)})\cdot\vec{s}}\left(\cos[(b_1 \vec{N}^{(1)}-b_2 \vec{N}^{(2)})\cdot\vec{t}] \right. \\ \nn
	 \left. + i \sin[(b_1 \vec{N}^{(1)}-b_2 \vec{N}^{(2)})\cdot\vec{t}] \right)
\eea
Solving these equations imposing $N_I^{(1)}=N_I^{(2)}$ and using the K\"ahler potential given by \eqref{newMthVx}, we find that the supersymmetric vacuum is given by:
\bea
 &&s_1 = \frac{\nu}{3N_1} \frac{1}{1-m} \qquad s_2 = \frac{\nu}{3N_2} \frac{1}{1-m} \qquad s_3 = \frac{\nu}{3N_3}\nn \\
 &&s_4 = \frac{2\nu}{3N_4} \frac{m}{1-m}\qquad s_5 = \frac{4\nu}{3N_5} 
\eea
The solution is very similar to \eqref{BobKanSusyVac}, apart for the combination $m=\frac{N_4^2}{2N_1N_2}$ appearing in the $s_1,s_2$ and $s_4$ moduli. We observed similar modification in the structure of the solutions for the flux vacua going from the K\"ahler potential of \cite{Acharya:2005ez} to ours \eqref{newMthVx}. We recall that here the $N_I^{(k)}$'s are not fluxes, but the numbers selecting the 3-cycle $Q_k$'s.

The same combination of the integers $N_I$ will presumably appear also in non-supersymmetric metastable vacua. These could present some relevant differences with respect to the results found in \cite{Acharya:2007rc,Acharya:2006ia}. It could be interesting to see how much they change.

\section{Conclusions}

In this paper we have studied a new ensemble of Type IIA flux vacua, in 
the CY with fluxes approximation. We have derived them by duality with 
M-theory flux vacua introduced in \cite{Acharya:2002kv}, where the author 
proved the moduli fixing in this setup.
In practise, we have considered a scalar potential for the moduli, where 
the superpotential is generated by form fluxes and gauge fluxes and a 
particular form of the K\"ahler potential $K$ is chosen. 
In particular, the superpotential contains a constant term which is non 
vanishing in the case of $G_2$ or Type IIA CY compactifications which 
admit a non zero complex Chern-Simons invariant. To our knowledge there is 
no obstruction to the existence of such compact manifolds (although a 
proof of existence is also missing). 
The obtained scalar potential describes moduli fixing both in Type IIA 
setup and in M-theory one, as the form is identical in the two cases; what 
changes is the interpretation of the solution. 

With this choice of the superpotential, the potential is simpler than that 
in \cite{DeWolfe:2005uu}. This allowed us to solve not only the 
supersymmetric equation $D_IW=0$, but even the equations $\partial_I V=0$. 
Actually we found all the extremal points, supersymmetric and 
non-supersymmetric. In particular we obtained that the absolute minimum is 
not supersymmetric.

The chosen form of the K\"ahler potential is suggested by studying 
toroidal orbifold compactifications. We have seen that, neglecting the 
twisted moduli, this form is quite general for about half of the orbifold 
models. We have also seen that it is consistent to neglect the twisted 
moduli: the physical quantities depend in fact mainly on the untwisted 
moduli. Finding the values the latter are fixed to is thus enough to study 
distributions of some important physical quantities, such as the classical 
cosmological constant or the gravitino mass. In Type IIA setup we have 
also seen that the tadpole cancellation condition does not give bound on 
fluxes, contrary to Type IIB flux vacua studied in \cite{Giddings:2001yu} 
and Type IIA flux vacua studied in \cite{DeWolfe:2005uu}. We derived a 
bound on fluxes by requiring  the validity of the supergravity 
approximation. To complete the study of orbifold models listed in Section 
\ref{TorOrb}, one should consider the case $h^{2,1}_{untw}=1$ and so study 
vacua coming from a new form of $K$. We do not expect a main difference in 
the results, being the form of the two K\"ahler potentials not so 
qualitative different.

The form of the K\"ahler potential we have used is also justified in 
M-theory by requiring it to satisfy the necessary conditions given by 
$G_2$ holonomy of the compactification manifold. These conditions were 
used in \cite{Acharya:2005ez} to guess the form of the K\"ahler potential 
$K=-3\ln V_X$ to be given by the $V_X$ in \eqref{VxMth}. This expression 
does not include our form of $K$, given by \eqref{newMthVx}. Actually, 
studying the M-theory vacua given by this new form of $K$ gives some 
differences in the distributions of the physical quantities. For example, 
in our ensemble of M-theory flux vacua it is easier to get small 
supersymmetry breaking scale than in the ensemble studied in 
\cite{Acharya:2005ez}. It would be interesting to find other forms of 
$V_X$ not included in \eqref{VxMth} and study moduli fixing and statistics 
of the resulting new vacua.

Another ensemble of M-theory vacua has been recently analysed in 
\cite{Acharya:2006ia,Acharya:2007rc}. It differs from 
\cite{Acharya:2005ez} in the superpotential, generated by non-perturbative 
effects, but not in the K\"ahler potential. It would be interesting to 
repeat their analysis, using the form of $K$ used in this present paper, 
and compare the results. Here we have briefly shown the supersymmetric 
minimum obtained using our form of the K\"ahler potential and we have seen 
that the difference with the previous result is in a particular 
combination of the parameters $N_I$. We have observed a similar difference 
between the generic flux vacua in studied in \cite{Acharya:2005ez} and 
those presented here. 

The study of the set of vacua presented in this paper is made through a 
four dimensional analysis as in \cite{DeWolfe:2005uu}: one derives the 
effective four dimensional potential and minimizes it. One can also try to 
solve directly the ten dimensional equations. In \cite{Acharya:2006ne} the 
authors followed this line and with a ten dimensional analysis they found 
the same results of \cite{DeWolfe:2005uu}. See also \cite{Banks:2006hg} 
for a ten dimensional approach to Type IIA and M-theory duals. It would be 
interesting to make similar studies for the set of vacua described here. 
This could also clarify the origin of the complex Chern-Simons term in 
Type IIA.

\vskip 1.5 cm

\begin{center} \textbf{Acknowledgements} \end{center}

We would like to thank Bobby Acharya, Qasem Exirifard, Martin O'Loughlin and Giuliano Panico for useful discussions. Particular thanks go to Martin O'Loughlin and Alexander Westphal for reading the draft and giving useful comments.

\newpage

\appendix
\section{Minimisation of the Potential}
In this section we discuss how to find the values of the stabilised moduli from the set of equations
\begin{equation}
 	\der_I V = 0\qquad I=1,\ldots,5
\end{equation}
The potential $V$ is given by
\begin{multline}
 	V= e^{K}\left( 4 K^{ij} N_i N_j + 4 c_2 W_2\right) = \\
=16 \bigg[\frac{\ell^2 N_5^2}4 + N_1^2 v_1^2 + N_4^2 v_1 v_2 + 
            N_2^2 v_2^2 + N_3^2 v_3^2 + 2 N_4 (N_1 v_1 + N_2 v_2) v_4 + 
            \frac 1 2 (2 N_1 N_2 + N_4^2) v_4^2 + \\+
            c_2 (c_2 + \ell N_5 + N_1 v_1 + N_2 v_2 + N_3 v_3 + 
                  N_4 v_4)\bigg]\left[\kappa \ell^4 (v_1 v_2 v_3 - v_3 \frac{v_4^2} 2)\right]^{-1}\,.
\end{multline}
The corresponding equations are
\begin{multline*}
 	\mathbf{E1}\equiv-v_2 [4 c_2^2 + \ell^2 N_5^2 - 4 N_1^2 v_1^2 + 4 N_2^2 v_2^2 + 
          4 N_3^2 v_3^2 + 4 c_2 (l N_5 + N_2 v_2 + N_3 v_3)]+\\ - 
    4 N_4 v_2 (c_2 + 2 N_2 v_2) v_4 - 
    2 [N_1 (c_2 + 2 N_1 v_1) + 2 (N_1 N_2 + N_4^2) v_2] v_4^2 - 
    4 N_1 N_4 v_4^3=0
\end{multline*}
\begin{multline*}
 	\mathbf{E2}\equiv-v_1 [4 c_2^2 + \ell^2 N_5^2 + 4 N_1^2 v_1^2 - 4 N_2^2 v_2^2 + 
          4 N_3^2 v_3^2 + 4 c_2 (l N_5 + N_1 v_1 + N_3 v_3)] +\\- 
    4 N_4 v_1 (c_2 + 2 N_1 v_1) v_4 - 
    2 (c_2 N_2 + 2 (N_1 N_2 v_1 + N_4^2 v_1 + N_2^2 v_2)) v_4^2 - 
    4 N_2 N_4 v_4^3=0
\end{multline*}
\begin{multline*}
 	\mathbf{E3}\equiv-4 c_2^2 - \ell^2 N_5^2 - 
    4 (N_1^2 v_1^2 + N_4^2 v_1 v_2 + N_2^2 v_2^2 - N_3^2 v_3^2) - 
    8 N_4 (N_1 v_1 + N_2 v_2) v_4+\\ - 2 (2 N_1 N_2 + N_4^2) v_4^2 - 
    4 c_2 (l N_5 + N_1 v_1 + N_2 v_2 + N_4 v_4)=0
\end{multline*}
\begin{multline*}
 	\mathbf{E4}\equiv4 N_4 v_1 v_2 (c_2 + 2 N_1 v_1 + 2 N_2 v_2) + (4 c_2^2 + 
          \ell^2 N_5^2 + 4 c_2 (l N_5 + N_1 v_1 + N_2 v_2 + N_3 v_3) +\\+ 
          4 (N_1^2 v_1^2 + 2 (N_1 N_2 + N_4^2) v_1 v_2 + 
                N_2^2 v_2^2 + N_3^2 v_3^2)) v_4 + 
    2 N_4 (c_2 + 2 N_1 v_1 + 2 N_2 v_2) v_4^2=0
\end{multline*}
\begin{multline*}
\mathbf{E5}\equiv-8 c_2^2 - \ell^2 N_5^2 - 
    8 (N_1^2 v_1^2 + N_4^2 v_1 v_2 + N_2^2 v_2^2 + N_3^2 v_3^2) - 
    16 N_4 (N_1 v_1 + N_2 v_2) v_4 +\\- 4 (2 N_1 N_2 + N_4^2) v_4^2 - 
    c_2 (6\ell N_5 + 8 (N_1 v_1 + N_2 v_2 + N_3 v_3 + N_4 v_4))=0
\end{multline*}
Combining $\mathbf{E1}$ with $\mathbf {E2}$ we get
\begin{equation}\label{v1E1v2E2}
 	v_1 \mathbf{E1} -v_2\mathbf{E2} \equiv(2 v_1v_2 - v_4^2)(N_1 v_1 - N_2 v_2)(c_2 + 
        2(N_1 v_1 + N_2 v_2 + N_4 v_4)=0
\end{equation}
The first factor is proportional to $\vol$ and thus cannot vanish.
We also combine $\mathbf{E5}$ with $\mathbf {E3}$ obtaining
\begin{equation}\label{b0b}
 	\mathbf{E5} - 2 \mathbf {E3}\equiv (l N_5 - 4 N_3 v_3) (2 c_2 +\ell N_5 + 4 N_3 v_3)\,.
\end{equation}
We have thus four possible branches of solutions specified by the choice of vanishing factors in \eqref{v1E1v2E2},\eqref{b0b}.
Let us consider the first branch:
\begin{gather}
N_1 v_1 - N_2 v_2 = 0\\
l N_5 - 4 N_3 v_3 = 0\,,
\end{gather}
solve these equations for $v_2,l$ and plug the solution into the remaining three equations. We can then solve the resulting system and get 11 solutions. Five of them have vanishing or imaginary moduli and only six are acceptable:
\begin{equation}
 	S_n = \big( v_1^{(n)}, v_3^{(n)}, v_4^{(n)})\qquad n = 1,\ldots,6
\end{equation}
with
{\tiny
\begin{equation*}
 S_1 = \left(\frac{2 c_2 N_2}{5 \left(-2 N_1 N_2+N_4^2\right)},-\frac{c_2}{5 N_3},\frac{2 c_2 N_4}{10 N_1 N_2-5 N_4^2}\right)
\end{equation*}
\begin{equation*}
 S_2 = \left(\frac{2 c_2 N_2}{-2 N_1 N_2+N_4^2},-\frac{c_2}{N_3},-\frac{2 c_2 N_4}{-2 N_1 N_2+N_4^2}\right)
\end{equation*}
\begin{multline*}
 	S_3=\left(-\frac{N_2 \left(2 c_2 N_1 N_2 N_4-c_2 N_4^3+\sqrt{6} \sqrt{c_2^2 N_4^2 \left(-2 N_1 N_2+N_4^2\right)^2}\right)}{5 N_4 \left(-2 N_1 N_2+N_4^2\right)^2},\right.\\\left.\frac{-4 c_2+\frac{\sqrt{6} \sqrt{c_2^2 N_4^2 \left(-2 N_1 N_2+N_4^2\right)^2}}{2 N_1 N_2 N_4-N_4^3}}{10 N_3},
\frac{2 c_2 N_1 N_2 N_4-c_2 N_4^3+\sqrt{6} \sqrt{c_2^2 N_4^2 \left(-2 N_1 N_2+N_4^2\right)^2}}{5 \left(-2 N_1 N_2+N_4^2\right)^2}\right)
\end{multline*}
\begin{multline*}
 	S_4=\left(\frac{N_2 \left(\sqrt{6} \sqrt{c_2^2 N_4^2 \left(-2 N_1 N_2+N_4^2\right)^2}+c_2 \left(-2 N_1 N_2 N_4+N_4^3\right)\right)}{5 N_4 \left(-2 N_1 N_2+N_4^2\right)^2},\right.\\\left.\frac{-4 c_2+\frac{\sqrt{6} \sqrt{c_2^2 N_4^2 \left(-2 N_1 N_2+N_4^2\right)^2}}{-2 N_1 N_2 N_4+N_4^3}}{10 N_3},-\frac{\sqrt{6} \sqrt{c_2^2 N_4^2 \left(-2 N_1 N_2+N_4^2\right)^2}+c_2 \left(-2 N_1 N_2 N_4+N_4^3\right)}{5 \left(-2 N_1 N_2+N_4^2\right)^2}\right)
\end{multline*}
\begin{multline*}
 S_5=\left(\frac{-10 c_2 N_1 N_2+N_4 \sqrt{130 c_2^2 N_1 N_2-\frac{40 \sqrt{10} \sqrt{c_2^4 N_1^2 N_2^2 \left(-2 N_1 N_2+N_4^2\right)^2}}{2 N_1 N_2-N_4^2}}}{20 N_1 \left(2 N_1 N_2-N_4^2\right)},\Big[10 c_2^3 N_1^2 N_2^2 \left(-2 N_1 N_2+N_4^2\right)+2 \sqrt{10} c_2 N_1 N_2 \sqrt{c_2^4 N_1^2 N_2^2 \left(-2 N_1 N_2+N_4^2\right)^2}+\right.\\\left. + c_2^2 N_1 N_2 N_4 \left(2 N_1 N_2-N_4^2\right) \sqrt{130 c_2^2 N_1 N_2-\frac{40 \sqrt{10} \sqrt{c_2^4 N_1^2 N_2^2 \left(-2 N_1 N_2+N_4^2\right)^2}}{2 N_1 N_2-N_4^2}}+\right.\\\left. -2 N_4 \sqrt{c_2^4 N_1^2 N_2^2 \left(-2 N_1 N_2+N_4^2\right)^2} \sqrt{13 c_2^2 N_1 N_2-\frac{4 \sqrt{10} \sqrt{c_2^4 N_1^2 N_2^2 \left(-2 N_1 N_2+N_4^2\right)^2}}{2 N_1 N_2-N_4^2}}\Big]\right.\\\left. \left[2 c_2 N_1 N_2 N_3 \left(2 N_1 N_2-N_4^2\right) \left(10 c_2 N_1 N_2-N_4 \sqrt{130 c_2^2 N_1 N_2-\frac{40 \sqrt{10} \sqrt{c_2^4 N_1^2 N_2^2 \left(-2 N_1 N_2+N_4^2\right)^2}}{2 N_1 N_2-N_4^2}}\right)\right]^{-1},\right.\\\left. \frac{5 c_2 N_4-\sqrt{130 c_2^2 N_1 N_2-\frac{40 \sqrt{10} \sqrt{c_2^4 N_1^2 N_2^2 \left(-2 N_1 N_2+N_4^2\right)^2}}{2 N_1 N_2-N_4^2}}}{20 N_1 N_2-10 N_4^2}\right)
\end{multline*}
\begin{multline*}
 	S_6=\left(-\frac{10 c_2 N_1 N_2+N_4 \sqrt{130 c_2^2 N_1 N_2-\frac{40 \sqrt{10} \sqrt{c_2^4 N_1^2 N_2^2 \left(-2 N_1 N_2+N_4^2\right)^2}}{2 N_1 N_2-N_4^2}}}{20 N_1 \left(2 N_1 N_2-N_4^2\right)},\Bigg[10 c_2^3 N_1^2 N_2^2 \left(-2 N_1 N_2+N_4^2\right)+2 \sqrt{10} c_2 N_1 N_2 \sqrt{c_2^4 N_1^2 N_2^2 \left(-2 N_1 N_2+N_4^2\right)^2}+\right.\\\left.+c_2^2 N_1 N_2 N_4 \left(-2 N_1 N_2+N_4^2\right) \sqrt{130 c_2^2 N_1 N_2-\frac{40 \sqrt{10} \sqrt{c_2^4 N_1^2 N_2^2 \left(-2 N_1 N_2+N_4^2\right)^2}}{2 N_1 N_2-N_4^2}}+\right.\\\left.+2 N_4 \sqrt{c_2^4 N_1^2 N_2^2 \left(-2 N_1 N_2+N_4^2\right)^2} \sqrt{13 c_2^2 N_1 N_2-\frac{4 \sqrt{10} \sqrt{c_2^4 N_1^2 N_2^2 \left(-2 N_1 N_2+N_4^2\right)^2}}{2 N_1 N_2-N_4^2}}\Bigg]\right.\\\left.\Bigg[2 c_2 N_1 N_2 N_3 \left(2 N_1 N_2-N_4^2\right) \left(10 c_2 N_1 N_2+N_4 \sqrt{130 c_2^2 N_1 N_2-\frac{40 \sqrt{10} \sqrt{c_2^4 N_1^2 N_2^2 \left(-2 N_1 N_2+N_4^2\right)^2}}{2 N_1 N_2-N_4^2}}\right)]\Bigg]^{-1},\right.\\\left.\frac{5 c_2 N_4+\sqrt{130 c_2^2 N_1 N_2-\frac{40 \sqrt{10} \sqrt{c_2^4 N_1^2 N_2^2 \left(-2 N_1 N_2+N_4^2\right)^2}}{2 N_1 N_2-N_4^2}}}{20 N_1 N_2-10 N_4^2}\right)
\end{multline*}}
Let us now consider the second branch
\begin{gather}
N_1 v_1 - N_2 v_2 = 0\\
2 c_2 + \ell N_5 + 4 N_3 v_3 = 0\,,
\end{gather}
solve again the equations for $v_2,l$ and plug the solutions in the remaining equations. In this case we have ten solutions out of which only five are physically acceptable.
{\tiny\begin{equation*}
 S_7=\left(\frac{4 c_2 N_2}{5 \left(-2 N_1 N_2+N_4^2\right)},-\frac{2 c_2}{5 N_3},\frac{4 c_2 N_4}{10 N_1 N_2-5 N_4^2}\right)
\end{equation*}
\begin{multline*}
 S_8=\left(\frac{N_2 \left(5 c_2 N_4 \left(-2 N_1 N_2+N_4^2\right)+\sqrt{10} \sqrt{c_2^2 N_4^2 \left(-2 N_1 N_2+N_4^2\right)^2}\right)}{5 N_4 \left(-2 N_1 N_2+N_4^2\right)^2},\frac{c_2^2 N_4 \left(-2 N_1 N_2+N_4^2\right)}{\sqrt{10} N_3 \sqrt{c_2^2 N_4^2 \left(-2 N_1 N_2+N_4^2\right)^2}},-\Bigg[5 c_2 N_4 \left(-2 N_1 N_2+N_4^2\right)+\right.\\\left.+\sqrt{10} \sqrt{c_2^2 N_4^2 \left(-2 N_1 N_2+N_4^2\right)^2}\Bigg]\Bigg[5 \left(-2 N_1 N_2+N_4^2\right)^2\Bigg]^{-1}\right)
\end{multline*}
\begin{multline*}
 S_9=\left(-\frac{N_2 \left(10 c_2 N_1 N_2 N_4-5 c_2 N_4^3+\sqrt{10} \sqrt{c_2^2 N_4^2 \left(-2 N_1 N_2+N_4^2\right)^2}\right)}{5 N_4 \left(-2 N_1 N_2+N_4^2\right)^2},-\frac{c_2^2 N_4 \left(-2 N_1 N_2+N_4^2\right)}{\sqrt{10} N_3 \sqrt{c_2^2 N_4^2 \left(-2 N_1 N_2+N_4^2\right)^2}},\Bigg[10 c_2 N_1 N_2 N_4-5 c_2 N_4^3+\right.\\\left.+\sqrt{10} \sqrt{c_2^2 N_4^2 \left(-2 N_1 N_2+N_4^2\right)^2}\Bigg]\Bigg[5 \left(-2 N_1 N_2+N_4^2\right)^2\Bigg]^{-1}\right)
\end{multline*}
\begin{multline*}
 S_{10} =\left(-\frac{10 c_2 N_1 N_2+N_4 \sqrt{66 c_2^2 N_1 N_2-\frac{24 \sqrt{6} \sqrt{c_2^4 N_1^2 N_2^2 \left(-2 N_1 N_2+N_4^2\right)^2}}{2 N_1 N_2-N_4^2}}}{20 N_1 \left(2 N_1 N_2-N_4^2\right)},\Bigg[10 c_2^3 N_1^2 N_2^2 \left(-2 N_1 N_2+N_4^2\right)-10 \sqrt{6} c_2 N_1 N_2 \sqrt{c_2^4 N_1^2 N_2^2 \left(-2 N_1 N_2+N_4^2\right)^2}+\right.\\\left.+c_2^2 N_1 N_2 N_4 \left(-2 N_1 N_2+N_4^2\right) \sqrt{66 c_2^2 N_1 N_2-\frac{24 \sqrt{6} \sqrt{c_2^4 N_1^2 N_2^2 \left(-2 N_1 N_2+N_4^2\right)^2}}{2 N_1 N_2-N_4^2}}+\right.\\\left.-6 N_4 \sqrt{c_2^4 N_1^2 N_2^2 \left(-2 N_1 N_2+N_4^2\right)^2} \sqrt{11 c_2^2 N_1 N_2-\frac{4 \sqrt{6} \sqrt{c_2^4 N_1^2 N_2^2 \left(-2 N_1 N_2+N_4^2\right)^2}}{2 N_1 N_2-N_4^2}}\Bigg]\right.\\\left.\Bigg[10 c_2 N_1 N_2 N_3 \left(2 N_1 N_2-N_4^2\right) \left(10 c_2 N_1 N_2+N_4 \sqrt{66 c_2^2 N_1 N_2-\frac{24 \sqrt{6} \sqrt{c_2^4 N_1^2 N_2^2 \left(-2 N_1 N_2+N_4^2\right)^2}}{2 N_1 N_2-N_4^2}}\right)\Bigg]^{-1},\right.\\\left.\frac{5 c_2 N_4+\sqrt{66 c_2^2 N_1 N_2-\frac{24 \sqrt{6} \sqrt{c_2^4 N_1^2 N_2^2 \left(-2 N_1 N_2+N_4^2\right)^2}}{2 N_1 N_2-N_4^2}}}{20 N_1 N_2-10 N_4^2}\right)
\end{multline*}
\begin{multline*}
 S_{11} =\left(\frac{-10 c_2 N_1 N_2+N_4 \sqrt{66 c_2^2 N_1 N_2-\frac{24 \sqrt{6} \sqrt{c_2^4 N_1^2 N_2^2 \left(-2 N_1 N_2+N_4^2\right)^2}}{2 N_1 N_2-N_4^2}}}{20 N_1 \left(2 N_1 N_2-N_4^2\right)},\Bigg[10 c_2^3 N_1^2 N_2^2 \left(-2 N_1 N_2+N_4^2\right)-10 \sqrt{6} c_2 N_1 N_2 \sqrt{c_2^4 N_1^2 N_2^2 \left(-2 N_1 N_2+N_4^2\right)^2}+\right.\\\left.+c_2^2 N_1 N_2 N_4 \left(2 N_1 N_2-N_4^2\right) \sqrt{66 c_2^2 N_1 N_2-\frac{24 \sqrt{6} \sqrt{c_2^4 N_1^2 N_2^2 \left(-2 N_1 N_2+N_4^2\right)^2}}{2 N_1 N_2-N_4^2}}+\right.\\\left.+6 N_4 \sqrt{c_2^4 N_1^2 N_2^2 \left(-2 N_1 N_2+N_4^2\right)^2} \sqrt{11 c_2^2 N_1 N_2-\frac{4 \sqrt{6} \sqrt{c_2^4 N_1^2 N_2^2 \left(-2 N_1 N_2+N_4^2\right)^2}}{2 N_1 N_2-N_4^2}}\Bigg]\right.\\\left.\Bigg[10 c_2 N_1 N_2 N_3 \left(2 N_1 N_2-N_4^2\right) \left(10 c_2 N_1 N_2-N_4 \sqrt{66 c_2^2 N_1 N_2-\frac{24 \sqrt{6} \sqrt{c_2^4 N_1^2 N_2^2 \left(-2 N_1 N_2+N_4^2\right)^2}}{2 N_1 N_2-N_4^2}}\right)]^{-1},\right.\\\left.\frac{5 c_2 N_4-\sqrt{66 c_2^2 N_1 N_2-\frac{24 \sqrt{6} \sqrt{c_2^4 N_1^2 N_2^2 \left(-2 N_1 N_2+N_4^2\right)^2}}{2 N_1 N_2-N_4^2}}}{20 N_1 N_2-10 N_4^2}\right)
\end{multline*}}

In the other two branches
\begin{gather}
 c_2 + 
        2(N_1 v_1 + N_2 v_2 + N_4 v_4)=0\\
\ell N_5 - 4 N_3 v_3 = 0
\end{gather}
and
\begin{gather}
 c_2 + 
        2(N_1 v_1 + N_2 v_2 + N_4 v_4)=0\\
2 c_2 + \ell N_5 + 4 N_3 v_3 = 0\,,
\end{gather}
two of the remaining three equations are linearly dependent and thus we do not have moduli fixing.
\section{Properties of Vacua and Constraints}
In this section we discuss the properties of the extremal point of the potential that we have obtained and the condition for them to be physical. We do not perform a consistency analysis like in the two cases of the supersymmetric vacuum and the absolute minimum both because the generic case is not as interesting as those and because of high computational complexity. 
As a first condition on the solutions, as already stated in Section \ref{modu_stab}, the kinetic term matrix $K_{IJ}$ must be positive definite. The eigenvalues of $K_{IJ}$ are given by
\bea
 &&\left\{ \frac 4 {\ell^2}\, ,\, \frac{1}{v_3^2} \, ,\, \frac{2\big(v_1^2+v_2^2+v_4^2-\sqrt{(v_1^2+v_2^2+v_4^2)^2-(v_4^2 -2 v_1 v_2)^2}\big)}{(v_4^2 -2 v_1 v_2)^2} \right. \\&&\left. \frac{2\big(v_1^2+v_2^2+v_4^2+\sqrt{(v_1^2+v_2^2+v_4^2)^2-(v_4^2 -2 v_1 v_2)^2}\big)}{(v_4^2 -2 v_1 v_2)^2} \, ,\, \frac{2}{2 v_1 v_2 - v_4^2}\right\}
\eea
and thus the metric is positive definite if and only if 
\begin{equation}
	 	2 v_1 v_2 - v_4^2 >0
\end{equation}
Looking at the solutions $S_n$ of the previous section, this is satisfied for
\begin{equation}\label{generic_const_K}
 	0<m<1 \quad \forall \,n
\end{equation}
The second requirement is the positiveness of the compactification volume. In string frame it is given by:
\begin{equation}
\vol = \kappa v_1 v_2 v_3  -\kappa v_3 v_4^2/2= \frac{\kappa v_3}{2} ( 2 v_1 v_2  - v_4^2)
\end{equation}
Taking into account the condition \eqref{generic_const_K}, this implies $\kappa v_3 >0$, which is satisfied for

\begin{equation}\label{c2N3}
 	\begin{cases}
 	 	\kappa c_2 N_3 < 0  & \mbox{for} \, n\neq 8,9\\
		\kappa N_4 N_3 < 0 & \mbox{for}	\, n=8\\
		\kappa N_4 N_3 > 0 & \mbox{for}	\, n=9
 	\end{cases}
\end{equation}
A third condition comes from the dilaton modulus $\ell$. It is related to the four dimensional dilaton $D$ and the ten dimensional dilaton $\phi$ as
\begin{equation}
 	\ell = \e^{-D} = \vol^{1/2} \e^{-\phi}
\end{equation}
which has to be a positive quantity. For all the extremal points $S_n$ this is equivalent to
\begin{equation}
 	c_2 N_5 <0\,.
\end{equation}
The next step is to evaluate the potential $V$ on the extremal points $S_n$ in order to distinguish different type of vacua and discuss their stability. We first summarise the results in a table and later discuss their derivation. 

For
\begin{equation}
 	\kappa c_2 N_3 <0 
\end{equation}
we get
\begin{equation}\label{tabledSAdSstabil}
 	\begin{array}{|c||cc|cc|}
\hline 
&\multicolumn{2}{c|}{c_2N_4<0}&\multicolumn{2}{c|}{c_2 N_4>0}\\
\hline
\hline
S_1& AdS&\textrm{St}&AdS&\textrm{St}\\
S_2 &dS&\textrm{Unst}&dS&\textrm{Unst}\\
S_3&dS&\textrm{Unst}&AdS&\textrm{St}\\
S_4&AdS&\textrm{St}&dS&\textrm{Unst}\\
S_5& AdS&\textrm{St}&AdS&\textrm{St}\\
S_6& AdS&\textrm{St}&AdS&\textrm{St}\\
\hline
S_7& AdS&\textrm{St}&AdS&\textrm{St}\\
S_8& \diagup&\diagup&AdS&\textrm{St}\\
S_9& AdS&\textrm{St}&\diagup&\diagup\\
S_{10}& AdS&\textrm{St}&AdS&\textrm{St}\\
S_{11}& AdS&\textrm{St}&AdS&\textrm{St}\\
\hline
\end{array}
\end{equation}
As noted in \eqref{c2N3} $S_{8,9}$ are the only two minima which are acceptable for $c_2 N_3 >0$ in this case they are $AdS$ but unstable for both signs of $c_2 N_4$.

The distinction between $dS$ and $AdS$ vacuum is easily determined by the sign on the potential evaluated on the solutions.
When the extremal point $S_n$ is a deSitter vacuum, it is stable if the matrix $\der_I\der_J V|_{S_n}$ is positive definite. 
On the other hand, an $AdS$ critical point $S_n$ does not have to be a local minimum to be perturbatively stable. It suffices that the eigenvalues of the Hessian of $V$ are not too negative compared to the cosmological constant, and more precisely that the Breitenlohner-Freedom bound [...] is satisfied:
\begin{equation}\label{stabeq}
 \frac{\der}{\der \hat{s}^H}\frac{\der}{\der \hat{s}^K}   V|_{S_n}-\frac{3}{2}V|_{S_n} \delta_{HK} > 0
\end{equation}
The derivatives are done with respect to the scalar with canonically normalised kinetic terms. This requires to diagonalise the K\"ahler metric and rescale the moduli. In fact the relevant kinetic term expanded around the critical point $S_n$ is (in four dimensional Planck units):
\begin{eqnarray}
g_{I\bar{J}} \der_\mu z^I \der^\mu \bar{z}^{\bar{J}} &=& \frac{1}{4} K_{IJ} \der_\mu s^I \der^\mu s^J + (\mbox{axions term}) \nn\\
    &=& \frac{1}{2} \delta_{HK} \der_\mu \hat{s}^H \der^\mu \hat{s}^K + (\mbox{axions term})
\end{eqnarray}
From here we get the relation:
\begin{equation}
\delta_{HK} = \frac{1}{2} K_{IJ}|_{S_n}{U^I}_K {U^J}_H
\end{equation}
where $U$ is a non singular matrix which depends on the extremal point $S_n$. The expression of the scalars $s^I$ in terms of canonically normalised scalars $\hat{s}^K$ is
\begin{equation}
s^I = {U^I}_J \hat{s}^J
\end{equation}
We find also the relation between the derivatives:
\begin{eqnarray}
\frac{\der}{\der s^I}= {V_I}^K \frac{\der}{\der \hat{s}^K} &\mbox{where}& {V_I}^K {U^J}_K = \delta_I^J
\end{eqnarray}
Multiplying \eqref{stabeq} by ${V_I}^K{V_J}^H$ and using the relations above, one gets the equivalent condition:
\begin{equation}
    \frac{\der}{\der s^I}\frac{\der}{\der s^J} V|_{S_n} - \frac 3 4 V|_{S_n} K_{IJ}|_{S_n} > 0 \:.
\end{equation}
A necessary and sufficient condition for a finite size matrix to be positive definite is that the determinants of all the diagonal are positive. Calculation of these determinants is not hard and allows us to draw the conclusions reported in table \eqref{tabledSAdSstabil}


\begin{thebibliography}{10}

\bibitem{Acharya:2005ez}
B.~S. Acharya, F.~Denef, and R.~Valandro, ``Statistics of M theory vacua,''
  {\em JHEP} {\bf 06} (2005) 056,
\href{http://arXiv.org/abs/hep-th/0502060}{{\tt hep-th/0502060}}.

\bibitem{Polchinski:1998rq}
J.~Polchinski, ``String theory. Vol. 1 and 2,''. Cambridge, UK: Univ. Pr.
  (1998).

\bibitem{Kaluza:1921tu}
T.~Kaluza, ``On the Problem of Unity in Physics,'' {\em Sitzungsber. Preuss.
  Akad. Wiss. Berlin (Math. Phys. )} {\bf 1921} (1921)
966--972.

\bibitem{Klein:1926tv}
O.~Klein, ``Quantum theory and five-dimensional theory of relativity,'' {\em Z.
  Phys.} {\bf 37} (1926)
895--906.

\bibitem{Grana:2005jc}
M.~Grana, ``Flux compactifications in string theory: A comprehensive review,''
  {\em Phys. Rept.} {\bf 423} (2006) 91--158,
\href{http://arXiv.org/abs/hep-th/0509003}{{\tt hep-th/0509003}}.

\bibitem{Douglas:2006es}
M.~R. Douglas and S.~Kachru, ``Flux compactification,'' {\em Rev. Mod. Phys.}
  {\bf 79} (2007) 733--796,
\href{http://arXiv.org/abs/hep-th/0610102}{{\tt hep-th/0610102}}.

\bibitem{Blumenhagen:2006ci}
R.~Blumenhagen, B.~Kors, D.~Lust, and S.~Stieberger, ``Four-dimensional String
  Compactifications with D-Branes, Orientifolds and Fluxes,''
\href{http://arXiv.org/abs/hep-th/0610327}{{\tt hep-th/0610327}}.

\bibitem{DeWolfe:2005uu}
O.~DeWolfe, A.~Giryavets, S.~Kachru, and W.~Taylor, ``Type IIA moduli
  stabilization,'' {\em JHEP} {\bf 07} (2005) 066,
\href{http://arXiv.org/abs/hep-th/0505160}{{\tt hep-th/0505160}}.

\bibitem{Beasley:2002db}
C.~Beasley and E.~Witten, ``A note on fluxes and superpotentials in M-theory
  compactifications on manifolds of G(2) holonomy,'' {\em JHEP} {\bf 07} (2002)
  046,
\href{http://arXiv.org/abs/hep-th/0203061}{{\tt hep-th/0203061}}.

\bibitem{Acharya:2002kv}
B.~S. Acharya, ``A moduli fixing mechanism in M theory,''
\href{http://arXiv.org/abs/hep-th/0212294}{{\tt hep-th/0212294}}.

\bibitem{Acharya:2000gb}
B.~S. Acharya, ``On realising N = 1 super Yang-Mills in M theory,''
\href{http://arXiv.org/abs/hep-th/0011089}{{\tt hep-th/0011089}}.

\bibitem{Acharya:2001gy}
B.~Acharya and E.~Witten, ``Chiral fermions from manifolds of G(2) holonomy,''
\href{http://arXiv.org/abs/hep-th/0109152}{{\tt hep-th/0109152}}.

\bibitem{Grimm:2004ua}
T.~W. Grimm and J.~Louis, ``The effective action of type IIA Calabi-Yau
  orientifolds,'' {\em Nucl. Phys.} {\bf B718} (2005) 153--202,
\href{http://arXiv.org/abs/hep-th/0412277}{{\tt hep-th/0412277}}.

\bibitem{Candelas:1990pi}
P.~Candelas and X.~de~la Ossa, ``MODULI SPACE OF CALABI-YAU MANIFOLDS,'' {\em
  Nucl. Phys.} {\bf B355} (1991)
455--481.

\bibitem{Kachru:2001je}
S.~Kachru and J.~McGreevy, ``M-theory on manifolds of G(2) holonomy and type
  IIA orientifolds,'' {\em JHEP} {\bf 06} (2001) 027,
\href{http://arXiv.org/abs/hep-th/0103223}{{\tt hep-th/0103223}}.

\bibitem{Breitenlohner:1982bm}
P.~Breitenlohner and D.~Z. Freedman, ``Positive Energy in anti-De Sitter
  Backgrounds and Gauged Extended Supergravity,'' {\em Phys. Lett.} {\bf B115}
  (1982)
197.

\bibitem{Ihl:2006pp}
M.~Ihl and T.~Wrase, ``Towards a realistic type IIA T**6/Z(4) orientifold model
  with background fluxes. I: Moduli stabilization,'' {\em JHEP} {\bf 07} (2006)
  027,
\href{http://arXiv.org/abs/hep-th/0604087}{{\tt hep-th/0604087}}.

\bibitem{Acharya:2007rc}
B.~S. Acharya, K.~Bobkov, G.~L. Kane, P.~Kumar, and J.~Shao, ``Explaining the
  electroweak scale and stabilizing moduli in M theory,''
\href{http://arXiv.org/abs/hep-th/0701034}{{\tt hep-th/0701034}}.

\bibitem{Acharya:2006ia}
B.~Acharya, K.~Bobkov, G.~Kane, P.~Kumar, and D.~Vaman, ``An M theory solution
  to the hierarchy problem,'' {\em Phys. Rev. Lett.} {\bf 97} (2006) 191601,
\href{http://arXiv.org/abs/hep-th/0606262}{{\tt hep-th/0606262}}.

\bibitem{Giddings:2001yu}
S.~B. Giddings, S.~Kachru, and J.~Polchinski, ``Hierarchies from fluxes in
  string compactifications,'' {\em Phys. Rev.} {\bf D66} (2002) 106006,
\href{http://arXiv.org/abs/hep-th/0105097}{{\tt hep-th/0105097}}.

\bibitem{Acharya:2006ne}
  B.~S.~Acharya, F.~Benini and R.~Valandro,
  ``Fixing moduli in exact type IIA flux vacua,''
  {\em JHEP} {\bf 0702} (2007) 018
  \href{http://arXiv.org/abs/hep-th/0607223}{{\tt hep-th/0607223}}.
\bibitem{Banks:2006hg}
  T.~Banks and K.~van den Broek,
  ``Massive IIA flux compactifications and U-dualities,''
  {\em JHEP} {\bf 0703}, 068 (2007)
  \href{http://arXiv.org/abs/hep-th/0611185}{{\tt hep-th/0611185}}.

\end{thebibliography}

\end{document}